\documentclass[11pt]{article}
\PassOptionsToPackage{hyphens}{url}

  \usepackage[final]{acl}

\usepackage[T1]{fontenc}
\usepackage[utf8]{inputenc}
\usepackage{times}
\usepackage{microtype}
\usepackage{amsmath,amssymb,amsthm,mathtools}
\usepackage{booktabs}
\usepackage{array}
\usepackage{multirow}
\usepackage{graphicx}
\usepackage{xcolor}
\usepackage{tikz}
\usepackage{pgfplots}
\pgfplotsset{compat=1.18}
\usepackage{enumitem}
\usepackage{url}
\usepackage{hyperref}

\hypersetup{hidelinks}
\newtheorem{theorem}{Theorem}
\newtheorem{lemma}[theorem]{Lemma}
\newtheorem{corollary}[theorem]{Corollary}
\newtheorem{proposition}[theorem]{Proposition}
\newtheorem{assumption}{Assumption}
\theoremstyle{definition}

\newcommand{\E}{\mathbb{E}}
\newcommand{\Prob}{\mathbb{P}}
\newcommand{\TV}{\operatorname{TV}}

\newcommand{\ind}{\mathbf{1}}

\newcommand{\Risk}{\mathcal{R}}
\newcommand{\FedLift}{\textsc{Fed-SRC}}
\newcommand{\FedContrast}{\textsc{Fed-SRC-C}}
\newcommand{\FedVA}{\textsc{Fed-SRC-VA}}
\newcommand{\eps}{\varepsilon}

\title{Private Anytime Selective-Risk Certification for Federated\\
Retrieval-Augmented Generation: Guarantees and Empirical Limits}

\author{
\textbf{Sanjeda Akter}\thanks{Equal contribution.}\textsuperscript{1}
\and
\textbf{Ibne Farabi Shihab}\footnotemark[1]\thanks{Corresponding author: \texttt{ishihab@iastate.edu}.}\textsuperscript{1}
\and
\textbf{Anuj Sharma}\textsuperscript{2}
\\[2pt]
\textsuperscript{1}Department of Computer Science, Iowa State University \\
\textsuperscript{2}Department of Civil, Construction \& Environmental Engineering, Iowa State University \\
\texttt{ishihab@iastate.edu}
}
\date{}

\begin{document}
\maketitle

\begin{abstract}
Selective-risk certificates promise that accepted outputs meet a declared error
target. We develop Fed-SRC, a score-agnostic certificate for federated,
differentially private, adaptively monitored retrieval-augmented generation.
Clients release only Gaussian-perturbed score and loss histograms.
Record-indexed and noise-variance-indexed martingales jointly bound target-risk
contrast and accepted mass over all registered thresholds and rounds, permitting
predictable recruitment, dropout, threshold selection, and optional stopping. A
range-one total-variation term transfers the calibration mixture to a declared
deployment mixture. The contribution is this private federated anytime
combination, not the contrast statistic or acceptance floor individually.
Empirically, no simultaneous-bound violation occurs in any cell, privacy level,
or policy. Operational power is score and population dependent: the primary
target $r_\star=0.10$ never certifies, and on RAGTruth the secondary target
$r_\star=0.20$ never certifies either, whereas on HaluEval question answering it
certifies in all 200 non-private trials with held-out risk below target. Naively
privatized non-private certificates violate their bounds in 146 to 198 of 200
trials. A private betting capital heuristic (we do not prove it is an e-process), run as an exploratory comparison
without proving its validity, stops certifying at $\eps\le4$ where Fed-SRC still
certifies. Certification still
consumes roughly 30 times more stream events than unique calibration items.
\end{abstract}

\section{Introduction}

Retrieval augmented language models can produce fluent claims unsupported by their evidence. Scores based on model confidence, sampling consistency, semantic entropy, or external verification can detect such errors, but they do not provide statistical guarantees \citep{manakul2023selfcheckgpt,farquhar2024semantic,min2023factscore}. Deployment instead requires a \emph{selective risk certificate}: accepted outputs must remain below a declared error target, with abstention when this cannot be established. Because this guarantee depends on the declared loss and target population, our certificate calibrates any bounded score rather than privileging one signal.

Information lift motivates this setting by comparing output probabilities with and without retrieved evidence \citep{akter2025infolift}. Large lift shows that evidence influenced generation, not that the output is correct. Under identical calibration on the same population, information lift never certifies on our benchmark, whereas a cheaper mechanistic detector does (\S\ref{sec:results}). This distinction motivates our focus on the certificate rather than the score.

Federated calibration introduces three challenges. Clients cannot pool prompts, outputs, or factuality labels; exact histograms can reveal membership; and continual threshold updates invalidate fixed time confidence intervals. Moreover, per example total variation does not imply a $BK$ bound for client blocks, empirical CDFs require explicit discretization and sensitivity, and repeated private releases require a simultaneous count scale noise envelope. \FedLift{} addresses these issues using finite histograms, explicit mixture transfer, and separate time uniform sampling and privacy noise martingales. Stable heterogeneity is absorbed into the realized calibration mixture, only the declared deployment mismatch $\eta_t$ is transferred, and privacy contributes its own count scale width.

\paragraph{Contributions.}
We formulate federated score calibration as selective risk control while separating the bounded ranking score from the declared loss. Two binned histograms, one for all examples and one weighted by loss, have add or remove $L_2$ sensitivity at most $\sqrt{2}$ regardless of the number of thresholds. We prove pathwise event level zCDP that charges each occurrence only for releases containing it. We also derive simultaneous range only and variance adaptive risk bounds with a joint acceptance floor, record indexed validity under predictable recruitment, and a realized noise envelope under dropout. A range one total variation argument transfers validity from the realized client mixture to the declared deployment mixture.

Empirically, we separate validity, power, and score quality. Across 500 trials for each core cell, the simultaneous bounds show no violation. The registered target $r_\star=0.10$ never certifies, while $r_\star=0.20$ certifies HaluEval question answering but not RAGTruth (\S\ref{sec:rq-reach}). At $r_\star=0.30$, information lift falls from 200/200 certifications without privacy to 2/200 at $\eps=4$, making the result valid but nearly vacuous at common privacy levels. Naive privatization violates validity and serves only as a stress test. On matched RAGTruth populations, information lift never certifies, whereas a mechanistic adaptation does, so a preregistered gate retires the lift centered claim. We therefore contribute and diagnose a private federated certificate without claiming a superior hallucination score or strong private utility on the current data.

\long\def\DetailedRelatedWork{%
\section{Expanded Related Work}

RAGTruth provides naturally generated RAG responses with response- and
span-level human annotations \citep{niu2024ragtruth}. SelfCheckGPT detects
sampling inconsistency \citep{manakul2023selfcheckgpt}, semantic entropy groups
generations by meaning before measuring uncertainty
\citep{farquhar2024semantic}, and ReDeEP uses mechanistic signals intended to
separate external-context use from parametric knowledge \citep{sun2024redeep}.
FRANQ is especially relevant because it separates faithfulness to retrieved
context from world factuality \citep{fadeeva2026franq}; ConU and SConU provide
conformalized self-consistency for open-ended generation
\citep{wang2024conu,wang2025sconu}. These methods are scores or prediction-set
baselines. Our certificate is score-agnostic and never pools faithfulness,
factuality, and answer correctness into one loss.

Information-lift certification and Sequential-EDFL motivate the score we set out
to certify \citep{akter2025infolift,akter2026sequential}. Sequential-EDFL certifies
a skeleton-relative generation-time information-lift property, not factual
correctness, federated privacy, or deployment selective risk. We do not claim a
new lift construction or a new anytime information-lift process.

Risk-controlling prediction sets, conformal risk control, and Learn-then-Test
give finite-sample calibration under explicit sampling assumptions
\citep{vovk2005algorithmic,bates2021risk,angelopoulos2024crc,
angelopoulos2025ltt}. C-RAG applies conformal risk analysis directly to RAG
generation and distribution shift \citep{kang2024crag}; our distinction is the
federated private adaptive transcript, not RAG risk certification itself.
Confidence sequences make inference uniform over time
\citep{howard2021time,ramdas2023game}. Anytime-Valid CRC extends CRC to a
growing calibration sample \citep{hultberg2026anytimecrc}. Conformal Selective
Acting already uses the gated excess-loss contrast for anytime LLM risk
\citep{khosravi2026csa}, and Yu and Liu give a variance-adaptive
selective-risk/acceptance certificate \citep{yu2026jointcertificate}. Their
methods are direct baselines; neither the contrast statistic nor an acceptance
floor is claimed as independently novel here.

Federated conformal methods address distributed quantiles, communication,
non-IID clients, and label shift
\citep{lu2023fcp,humbert2023oneshot,plassier2023labelshift}. Anytime-FC-RAG
provides federated anytime coverage while leaving DP outside its first version
\citep{dubey2026anytimefcrag}. Private confidence sequences and private
e-values show that privacy and anytime validity can coexist
\citep{waudbysmith2023privatecs,csillag2025dpevalues,
jacobsen2026dpeprocess}; recent private conformal work studies efficient
private calibration \citep{wu2026dpcp}. Non-exchangeable conformal work also
motivates attaching total-variation terms to observations or expectations,
not treating one per-example distance as one block-level penalty
\citep{barber2023beyond}.

Accordingly, ``anytime,'' ``selective LLM risk,'' ``federated,'' ``private
anytime inference,'' the gated contrast, and variance-adaptive selective-risk
certificates are all prior art. The contribution evaluated here is their joint
deployment contract: a finite event-level-DP federated release, a pathwise
accountant, validity under certificate-driven recruitment and stopping, and
transfer from the adaptively realized client mixture to a declared deployment
mixture.

}

\section{Closest Work}

RAG faithfulness and uncertainty scores include SelfCheckGPT, semantic
entropy, ReDeEP, FRANQ, ConU, and SConU
\citep{manakul2023selfcheckgpt,farquhar2024semantic,sun2024redeep,
fadeeva2026franq,wang2024conu,wang2025sconu}; information-lift certification
and Sequential-EDFL motivate our score but certify neither factual
correctness nor federated private deployment risk
\citep{akter2025infolift,akter2026sequential}, and C-RAG establishes
conformal RAG risk certification \citep{kang2024crag}.  Every statistical
component also has close precedent: CRC/LTT and confidence sequences
\citep{angelopoulos2024crc,angelopoulos2025ltt,howard2021time}; Anytime-Valid
CRC, CSA, and the Yu and Liu certificate for growing calibration, the gated
contrast, and variance-adaptive risk/acceptance
\citep{hultberg2026anytimecrc,khosravi2026csa,yu2026jointcertificate};
federated conformal prediction, Anytime-FC-RAG, private confidence sequences,
and private e-values
\citep{lu2023fcp,humbert2023oneshot,dubey2026anytimefcrag,
waudbysmith2023privatecs,csillag2025dpevalues}.  We claim none of these in
isolation; the contribution is their event-level-DP federated anytime
combination, pathwise accountant, and declared-mixture transfer.  The
expanded comparison is in the appendix.

\section{Problem Setup}
\label{sec:setup}

\subsection{Evidence lift is a score, not a truth label}

An example is
$W=(q,z,y,\ell)$: query $q$, retrieved evidence $z$, model output
$y=(y_1,\ldots,y_L)$, and bounded loss $\ell\in[0,1]$.  For response-level
faithfulness, $\ell=1$ can denote that at least one atomic claim is unsupported;
for a graded endpoint, $\ell$ may be the fraction of unsupported atomic claims.
The theory applies to either choice.  It certifies risk relative to this
declared annotation rule.

Let $p_\theta$ be a probability-exposed language model and $z^\varnothing$ the
same prompt with retrieved evidence removed while all other prompt fields are
held fixed.  For a clipping level $B>0$, define normalized evidence lift
\begin{equation}
S_B(W)
=
\frac{1}{LB}
\sum_{r=1}^{L}
\left[
\log
\frac{
p_\theta(y_r\mid y_{<r},q,z)
}{
p_\theta(y_r\mid y_{<r},q,z^\varnothing)
}
\right]_{[0,B]},
\label{eq:lift}
\end{equation}
which lies in $[0,1]$.
Here $[u]_{[a,b]}=\min\{b,\max\{a,u\}\}$.
Teacher forcing evaluates both probabilities on the same realized output.
Other fixed, bounded transformations of information lift are allowed.  The
score function, model, prompts, clipping level, and threshold grid must be fixed
before using the calibration records, or selected on a separate split.

Choose thresholds
$0\le\lambda_1<\cdots<\lambda_m\le1$ and define
\begin{align}
A_j(W)&=\ind\{S_B(W)\ge\lambda_j\},
\label{eq:accept}\\
Z_j(W)&=\ell\,A_j(W).
\label{eq:errorcount}
\end{align}
For a predeclared target $r_\star\in(0,1)$, also define the gated
target-risk contrast
\begin{equation}
D_j(W)=Z_j(W)-r_\star A_j(W)=A_j(W)(\ell-r_\star),
\label{eq:contrast}
\end{equation}
which lies in $[-r_\star,1-r_\star]$.
For a distribution $Q$, write
\begin{align}
a_{Q,j}&=\E_Q[A_j(W)],&
z_{Q,j}&=\E_Q[Z_j(W)],\notag\\
d_{Q,j}&=\E_Q[D_j(W)],&
\Risk_{Q,j}&=\frac{z_{Q,j}}{a_{Q,j}},
\label{eq:risk}
\end{align}
the last when $a_{Q,j}>0$.
Thus $a_{Q,j}$ is acceptance and $\Risk_{Q,j}$ is expected loss conditional on
acceptance.  If no output is accepted, we make no risk claim and return the
always-abstain policy. When $a_{Q,j}>0$, $\Risk_{Q,j}\le r_\star$ if and only
if $d_{Q,j}\le0$. Operational certification additionally requires a
predeclared floor $a_{\min}>0$; the contrast inequality alone is not called a
certificate.

\subsection{Federated calibration and the deployment target}

There are $K$ clients with calibration laws $P_k$, which may be arbitrarily heterogeneous. Index released calibration records by their order of inclusion $i=1,2,\ldots$; before record $i$ is revealed the scheduler picks its client $C_i\in[K]$ using only the past, and then $W_i\sim P_{C_i}$. This record-indexed form is what lets the next client and batch size depend on an earlier private certificate. Round $t$ is the prefix ending at $N_t$, with
\begin{align}
N_{k,t}&=\sum_{i=1}^{N_t}\ind\{C_i=k\},&
N_t&=\sum_{k=1}^{K}N_{k,t},\notag\\
\bar w_{k,t}&=N_{k,t}/N_t,
\label{eq:weights}
\end{align}
and we monitor only $\mathcal T_+=\{t:N_t\ge1\}$. Client identity, release status, and padded batch capacity are public metadata, and each successful message uses its announced number of contribution slots, so $N_{k,t}$ and $N_t$ are unchanged between neighbouring datasets. Our adjacency is the zero-out relation on one stream slot: one event occurrence contributes its histogram vector or zero while message shape remains fixed. Repeated events from the same person or source item are distinct slots and require composition for person- or source-level privacy. Hiding participation counts would need a private denominator and is outside this protocol. The round-$t$ calibration mixture is
\begin{equation}
\bar Q_t=\sum_{k=1}^{K}\bar w_{k,t}P_k.
\label{eq:calmix}
\end{equation}

The intended deployment mixture is
$Q^\star=\sum_{k=1}^{K}w_kP_k^\star$, where $w$ is a declared vector of
nonnegative weights summing to one and $P_k^\star$ is client $k$'s deployment
law.  We use the convention
$\TV(P,Q)=\sup_D|P(D)-Q(D)|$.  Suppose an independently justified bound
$\TV(P_k^\star,P_k)\le\gamma_k$ is available.  Define
\begin{equation}
\eta_t
=
\frac12\sum_{k=1}^{K}|w_k-\bar w_{k,t}|
+
\sum_{k=1}^{K}w_k\gamma_k.
\label{eq:eta}
\end{equation}
The first term is known from participation counts and target weights.  The
second is a declared sensitivity parameter unless drift is bounded from an
independent audit sample.  When deployment uses the realized calibration
mixture and components are stable, $\eta_t=0$.

\subsection{Desired guarantee}

Given target risk $r_\star\in(0,1)$, the server may inspect every private
transcript, choose any threshold at any round, and stop at a data-dependent
time.  We seek an upper certificate $U_{j,t}$ such that
\begin{equation}
\Prob\!\left[
\forall t\in\mathcal T_+,\ \forall j\in[m]:
\Risk_{Q^\star,j}\le U_{j,t}
\right]
\label{eq:goal}
\end{equation}
is at least $1-\alpha_{\rm s}-\alpha_{\rm n}$.
Here $\alpha_{\rm s}$ is the sampling-error budget and $\alpha_{\rm n}$ the
Gaussian-noise budget.  A selected threshold is certified if
$U_{j,t}\le r_\star$.

\section{\FedLift: Private Federated Calibration}
\label{sec:method}

\subsection{A finite sufficient release}

The thresholds partition $[0,1]$ into $m+1$ bins:
\begin{align}
I_0&=[0,\lambda_1),\\
I_b&=[\lambda_b,\lambda_{b+1})\quad (1\le b<m),\\
I_m&=[\lambda_m,1].
\end{align}
Let $\mathcal R_t\subseteq[K]$ be the clients that actually release in round
$t$; a client that drops out before releasing contributes neither a message nor
records to $N_t$.  For $(k,t)$ with $k\in\mathcal R_t$, define score and error
histograms
\begin{align}
C_{k,t,b}
&=\sum_{W\in D_{k,t}}\ind\{S_B(W)\in I_b\},
\label{eq:count-hist}\\
E_{k,t,b}
&=\sum_{W\in D_{k,t}}\ell(W)\,
  \ind\{S_B(W)\in I_b\}.
\label{eq:error-hist}
\end{align}
The client releases the $2(m+1)$-vector
\begin{equation}
\widetilde H_{k,t}
=
(C_{k,t,0:m},E_{k,t,0:m})+\xi_{k,t},
\label{eq:release}
\end{equation}
with $\xi_{k,t}\sim\mathcal N(0,\sigma_{k,t}^2I)$.
No prompt, evidence, output, per-example score, or per-example label is sent.

For threshold $j$, the server obtains noisy cumulative accepted and error
counts by summing bins $b=j,\ldots,m$, clients, and rounds:
\begin{align}
\widetilde A_{j,t}
&=
\sum_{s\le t}\sum_{k\in\mathcal R_s}\sum_{b=j}^{m}
\widetilde C_{k,s,b},
\label{eq:noisy-A}\\
\widetilde Z_{j,t}
&=
\sum_{s\le t}\sum_{k\in\mathcal R_s}\sum_{b=j}^{m}
\widetilde E_{k,s,b}.
\label{eq:noisy-Z}
\end{align}
Let $d_j=m-j+1$ be the number of suffix bins and
\begin{equation}
V_{j,t}
=
d_j\sum_{s\le t}\sum_{k\in\mathcal R_s}\sigma_{k,s}^2.
\label{eq:noisevar}
\end{equation}
Thus $V_{j,t}$ uses realized releases, not an all-client-per-round idealization.
Conditional on the past, each suffix-noise increment is centered Gaussian with
its corresponding variance increment.  Suffix sums are correlated across
thresholds, but the proof never assumes independence across $(j,t)$.

\subsection{Range-only and variance-adaptive contrast certificates}
\label{sec:contrast}

Let $\pi_r=6/[\pi^2(r+1)^2]$. For $c\ge1$, $n\ge1$, and $v\ge0$, define
\begin{align}
H_c(n;\alpha)
&=\sqrt{\frac{2^{\lceil\log_2n\rceil}}{2}
\log\frac{cm}{\alpha\pi_{\lceil\log_2n\rceil}}},
\label{eq:H}\\
G_c(v;\alpha)
&=\begin{cases}
0,&v=0,\\
\sqrt{2u_v\log\dfrac{cm}{\alpha\pi_{r_v}}},&v>0,
\end{cases}
\label{eq:G}
\end{align}
where $r_v=\max\{0,\lceil\log_2(v/v_0)\rceil\}$,
$u_v=v_0 2^{r_v}$, and $v_0>0$ is fixed before calibration. The
factor $c$ records how many one-sided process families share a probability
budget.

Write $V^A_{j,t}=V_{j,t}$,
$V^D_{j,t}=(1+r_\star^2)V^A_{j,t}$, and
$\widetilde D_{j,t}=\widetilde Z_{j,t}-r_\star\widetilde A_{j,t}$.
The range-only construction is
\begin{align}
\overline d^{\rm H}_{j,t}
&=\min\Bigl\{1-r_\star,\ \tfrac{1}{N_t}\bigl(\widetilde D_{j,t}+H_2(N_t;\alpha_{\rm s})\notag\\
&\qquad+G_2(V^D_{j,t};\alpha_{\rm n})\bigr)+\eta_t\Bigr\},
\label{eq:dH}\\
\underline a^{\rm H}_{j,t}
&=\max\Bigl\{0,\ \tfrac{1}{N_t}\bigl(\widetilde A_{j,t}-H_2(N_t;\alpha_{\rm s})\notag\\
&\qquad-G_2(V^A_{j,t};\alpha_{\rm n})\bigr)-\eta_t\Bigr\}.
\label{eq:aH}
\end{align}
It pays one sampling width, one Gaussian width, and one mixture-transfer term
for the target-risk decision. The separate risk-ratio certificate is retained
as a baseline in Appendix~\ref{app:ratiocert}.

For a variance-adaptive alternative, note that $D_j(W)^2\le A_j(W)$, so the
predictable quadratic variation of the contrast martingale is bounded by
expected accepted mass. Define
\begin{align}
\overline q_{j,t}
&=\min\Bigl\{N_t,\max\bigl\{0,\widetilde A_{j,t}+H_3(N_t;\alpha_{\rm s})\notag\\
&\qquad+G_3(V^A_{j,t};\alpha_{\rm n})\bigr\}\Bigr\},
\label{eq:qbar}\\
F_3(q;\alpha)
&=\max_{0\le r\le R_q}
\bigl\{\sqrt{2^{r+1}x_r}+\tfrac{2}{3}x_r\bigr\},
\label{eq:F}
\end{align}
where $R_q=\lceil\log_2\max\{q,1\}\rceil$ and
$x_r=\log\{3m/(\alpha\pi_r)\}$. Then
\begin{align}
\overline d^{\rm VA}_{j,t}
&=\min\Bigl\{1-r_\star,\ \tfrac{1}{N_t}\bigl(\widetilde D_{j,t}+F_3(\overline q_{j,t};\alpha_{\rm s})\notag\\
&\qquad+G_3(V^D_{j,t};\alpha_{\rm n})\bigr)+\eta_t\Bigr\},
\label{eq:dVA}\\
\underline a^{\rm VA}_{j,t}
&=\max\Bigl\{0,\ \tfrac{1}{N_t}\bigl(\widetilde A_{j,t}-H_3(N_t;\alpha_{\rm s})\notag\\
&\qquad-G_3(V^A_{j,t};\alpha_{\rm n})\bigr)-\eta_t\Bigr\}.
\label{eq:aVA}
\end{align}
The range-only and variance-adaptive constructions each receive the declared
budgets in a separately preregistered run. Selecting the narrower one after
seeing the data requires adding the construction index to the simultaneous
family or splitting the budgets.

\begin{theorem}[Private federated anytime target-risk certificate]
\label{thm:contrast}
Suppose client identity, batch size, release decision, threshold family, score,
loss, privacy scale, and each monitored prefix are predictable, and fresh
records and Gaussian noise follow their declared conditional laws. For either
$M\in\{{\rm H},{\rm VA}\}$, with probability at least
$1-\alpha_{\rm s}-\alpha_{\rm n}$,
\begin{equation}
d_{Q^\star,j}\le\overline d^M_{j,t},\qquad
a_{Q^\star,j}\ge\underline a^M_{j,t}
\label{eq:joint}
\end{equation}
simultaneously for every registered threshold $j$ and monitored round $t$.
Therefore any transcript-measurable threshold and stopping time satisfying
\begin{equation}
\overline d^M_{\widehat j,\widehat t}\le0,
\qquad
\underline a^M_{\widehat j,\widehat t}\ge a_{\min}>0
\label{eq:cert-rule}
\end{equation}
obeys $\Risk_{Q^\star,\widehat j}\le r_\star$ and
$a_{Q^\star,\widehat j}\ge a_{\min}$ on the same event. No additional
optional-stopping or post-selection correction is needed.
\end{theorem}

The direct transfer in the proof applies total variation to
$D=A(\ell-r_\star)\in[-r_\star,1-r_\star]$, whose range length is one. It
therefore pays $\eta_t$, not $(1+r_\star)\eta_t$. We make no universal
dominance claim over the ratio rule after clipping and stitching; matched
comparisons are empirical.

\subsection{Threshold selection}

At any round, the server forms
\begin{equation}
\mathcal{J}^M_t=\{j\in[m]:
\overline d^M_{j,t}\le0,
\ \underline a^M_{j,t}\ge a_{\min}\}.
\end{equation}
If $\mathcal{J}^M_t=\varnothing$, it returns the always-abstain policy. Otherwise it
selects the candidate with the largest conservative acceptance lower bound:
\begin{equation}
\widehat j_t\in\arg\max_{j\in\mathcal{J}^M_t}\underline a^M_{j,t}.
\label{eq:selection}
\end{equation}
Any other transcript-measurable choice is also valid.  Post-processing the
private transcript incurs no additional privacy loss.  This is not an LTT
step: the single simultaneous event in Theorem~\ref{thm:contrast} already covers
every registered threshold, so selecting among those bounds is direct
post-processing rather than a separate fixed-time familywise test.

\paragraph{Protocol summary.}
\begin{enumerate}[leftmargin=*,itemsep=2pt]
    \item Fix the model, lift transformation, loss, threshold grid, privacy
    level, risk target, deployment weights, and $v_0$ before calibration.
    \item Each client bins each new record once and releases the noised pair of
    histograms in Equation~\eqref{eq:release}.
    \item The server aggregates suffix counts, computes either
    Equations~\eqref{eq:dH} and \eqref{eq:aH} or
    Equations~\eqref{eq:dVA} and \eqref{eq:aVA} whenever $N_t\ge1$, and may
    inspect them after every round.
    \item Deploy the selected threshold only if both inequalities in
    Equation~\eqref{eq:cert-rule} hold; otherwise abstain.
\end{enumerate}

\long\def\DetailedGuarantees{%
\section{Full Guarantee Statements}
\label{sec:theory-full}

\begin{assumption}[Calibration stream]
\label{ass:stream}
Let $\mathcal F_{i-1}$ contain all calibration records and release randomness
strictly before record $i$.  The identity $C_i$ is
$\mathcal F_{i-1}$-measurable and
$W_i\mid\mathcal F_{i-1}\sim P_{C_i}$.  Monitored round endpoints are
non-anticipating prefixes.  Each realized release set, batch size, and noise
scale is fixed before drawing its fresh data and Gaussian noise; each fresh
noise vector has the conditional distribution in
Equation~\eqref{eq:release}.  The score, loss, clipping level, bins, and model
are fixed before these records are observed.  For the privacy statement, every
server-visible scheduling action is a function only of public metadata,
external randomness, and previous DP transcript messages.
\end{assumption}

Assumption~\ref{ass:stream} allows different laws for every client and permits
certificate-driven recruitment, escalation, and stopping: the next action may
depend on the entire earlier private transcript.  It excludes choosing a
client or retaining a record after inspecting that same unseen record.  It
also excludes training the lift score on the certification records; that case
requires a separate split or another uniform-validity argument.

\subsection{Privacy}

\begin{lemma}[Histogram sensitivity]
\label{lem:sensitivity}
Under add/remove record adjacency and $\ell\in[0,1]$, the concatenated count
and error histogram in Equations~\eqref{eq:count-hist} and \eqref{eq:error-hist}
has $L_2$ sensitivity at most $\sqrt2$.  The bound is attained when the added
record has $\ell=1$.
\end{lemma}

\begin{theorem}[Transcript privacy]
\label{thm:privacy}
For each message, Equation~\eqref{eq:release} is
$\rho_{k,t}$-zCDP with
\begin{equation}
\rho_{k,t}=\frac{1}{\sigma_{k,t}^{2}}.
\label{eq:rho-message}
\end{equation}
For any calibration record $i$ and admissible transcript path $\tau$, let
$\mathcal I_\tau(i)$ be the releases to which that record contributes.  If a
deterministic privacy filter enforces the pathwise bound
\begin{equation}
\rho_{\rm tr}
=
\sup_{\tau}\sup_i
\sum_{(k,t)\in\mathcal I_\tau(i)}
\frac{1}{\sigma_{k,t}^{2}}
\label{eq:rho-transcript}
\end{equation}
then the complete adaptive transcript is $\rho_{\rm tr}$-zCDP.
Consequently, for every $\delta_{\rm priv}\in(0,1)$, it is
$(\eps_{\rm priv},\delta_{\rm priv})$-DP with
\begin{equation}
\eps_{\rm priv}
=
\rho_{\rm tr}
+
2\sqrt{
\rho_{\rm tr}\log(1/\delta_{\rm priv})
}.
\label{eq:zcdp-conversion}
\end{equation}
In the disjoint-batch protocol, each record appears in at most one release and
$\rho_{\rm tr}=\sup_\tau\max_{(k,t)\in\tau}\sigma_{k,t}^{-2}$; the number of
clients and rounds does not enter that event's privacy loss.  If one person or
source record contributes several stream events, all corresponding releases
must instead be included in $\mathcal I_\tau(i)$. For a fixed
nonadaptive schedule, the suprema reduce to the corresponding realized maxima.
\end{theorem}

For a target $(\eps_{\rm priv},\delta_{\rm priv})$, define
\begin{equation}
\rho_\star
=
\left(
\sqrt{\log(1/\delta_{\rm priv})+\eps_{\rm priv}}
-
\sqrt{\log(1/\delta_{\rm priv})}
\right)^2.
\label{eq:rho-target}
\end{equation}
Using $\sigma_{k,t}\ge1/\sqrt{\rho_\star}$ for disjoint batches is sufficient.
This calibration is valid for all $\eps_{\rm priv}>0$.  By contrast, the
classical sufficient calibration
$\sigma\ge\Delta_2\sqrt{2\log(1.25/\delta)}/\eps$ is stated for
$\eps\in(0,1)$ \citep{dwork2014algorithmic}.  The analytic Gaussian mechanism
can calibrate a single fixed release more tightly \citep{balle2018analytic}; we
use zCDP because Equation~\eqref{eq:rho-transcript} composes transparently over
the adaptive transcript, including the experimental values $\eps>1$.

\subsection{Anytime-valid selective risk}

\begin{lemma}[Simultaneous clean-count bounds]
\label{lem:sampling}
Under Assumption~\ref{ass:stream}, with probability at least
$1-\alpha_{\rm s}$, simultaneously for every $t\in\mathcal T_+$ and
$j\in[m]$,
\begin{align}
N_tz_{\bar Q_t,j}
&\le Z_{j,t}+h_t(\alpha_{\rm s}),
\label{eq:sample-z}\\
N_ta_{\bar Q_t,j}
&\ge A_{j,t}-h_t(\alpha_{\rm s}),
\label{eq:sample-a}
\end{align}
where $A_{j,t}$ and $Z_{j,t}$ are the corresponding unnoised cumulative
counts.
\end{lemma}

\begin{lemma}[Simultaneous Gaussian envelopes]
\label{lem:noise}
With probability at least $1-\alpha_{\rm n}$, simultaneously for every
$t\in\mathcal T_+$ and $j\in[m]$,
\begin{align}
Z_{j,t}
&\le\widetilde Z_{j,t}+g_{j,t}(\alpha_{\rm n}),
\label{eq:noise-z}\\
A_{j,t}
&\ge\widetilde A_{j,t}-g_{j,t}(\alpha_{\rm n}).
\label{eq:noise-a}
\end{align}
\end{lemma}

\begin{lemma}[Client-mixture transfer]
\label{lem:shift}
For $Q^\star$, $\bar Q_t$, and $\eta_t$ defined in
Equations~\eqref{eq:calmix} and \eqref{eq:eta},
\begin{equation}
\TV(Q^\star,\bar Q_t)\le\eta_t.
\label{eq:tv-mixture}
\end{equation}
Hence for every $j$,
\begin{align}
z_{Q^\star,j}&\le z_{\bar Q_t,j}+\eta_t,\label{eq:zshift}\\
a_{Q^\star,j}&\ge a_{\bar Q_t,j}-\eta_t.\label{eq:ashift}
\end{align}
\end{lemma}

\begin{theorem}[Private federated anytime risk certificate]
\label{thm:main}
Under Assumption~\ref{ass:stream}, with probability at least
$1-\alpha_{\rm s}-\alpha_{\rm n}$ over the calibration records and DP noise,
\begin{equation}
\Risk_{Q^\star,j}\le U_{j,t}
\quad
\text{for all }t\in\mathcal T_+\text{ and }j\in[m]
\label{eq:main-simultaneous}
\end{equation}
whenever the denominator in Equation~\eqref{eq:risk} is positive.
Therefore, for any stopping time $\widehat t$ and any threshold
$\widehat j$ chosen as an arbitrary measurable function of the private
transcript up to $\widehat t$,
\begin{equation}
U_{\widehat j,\widehat t}\le r_\star
\quad\Longrightarrow\quad
\Risk_{Q^\star,\widehat j}\le r_\star
\label{eq:selected-guarantee}
\end{equation}
on the same event.  No additional multiple-testing or optional-stopping
correction is required.
\end{theorem}

\begin{corollary}[Width decomposition]
\label{cor:width}
Ignoring clipping at $0$ and $1$, for $V_{j,t}>0$ the error and acceptance
mean bounds each contain
\begin{align}
&\underbrace{
O\!\left(
\sqrt{
\frac{
\log(m/\alpha_{\rm s})
+\log\log(eN_t)
}{N_t}
}
\right)
}_{\text{sampling}}
\notag\\[-2pt]
&\quad+
\underbrace{
O\!\left(
\frac{\sqrt{\max\{V_{j,t},v_0\}}}{N_t}
\sqrt{
\log(m/\alpha_{\rm n})
+\log\log\!\left(
e+\frac{V_{j,t}}{v_0}
\right)
}
\right)
}_{\substack{\text{privacy}\\\text{noise}}}.
\label{eq:width}
\end{align}
and the noise term is zero when $V_{j,t}=0$.  The risk
transfer additionally pays $\eta_t$ in both numerator and denominator.  With
$K$ equal-noise clients each contributing $n$ records in
each of $t$ rounds and $v_0\le d_jKt\sigma^2$,
$N_t=Knt$ and $V_{j,t}=d_jKt\sigma^2$, so the privacy term scales as
\begin{equation}
\frac{\sigma}{n\sqrt{Kt}}\,
O\!\left(
\sqrt{d_j\bigl\{\log(m/\alpha_{\rm n})+
\log\log(e+d_jKt\sigma^2/v_0)\bigr\}}
\right).
\label{eq:balanced-width}
\end{equation}
\end{corollary}

\paragraph{Interpretation.}
The theorem separates three distinct issues:
\begin{itemize}[leftmargin=*,itemsep=1pt]
    \item $\alpha_{\rm s}+\alpha_{\rm n}$ is the probability that the
    simultaneous certificate fails.
    \item $\sigma$ is chosen by the privacy target and widens the certificate.
    \item $\eta_t$ encodes deployment-mixture mismatch and within-client drift.
\end{itemize}
Unlike an additive ``$\alpha+BK+\rho$'' statement, every quantity is either an
auditable probability budget or a width/shift term with explicit units and
constants.

}

\section{Guarantees}
\label{sec:theory}

Adding one bounded-loss record changes one count coordinate by one and one
loss coordinate by at most one, so the released vector has $L_2$ sensitivity
at most $\sqrt2$. A release with coordinate noise variance $\sigma^2$ is
$1/\sigma^2$-zCDP. A deterministic pathwise filter composes only the releases
in which a stream occurrence participates and converts the resulting $\rho$ to
$(\rho+2\sqrt{\rho\log(1/\delta)},\delta)$-DP. The complete adaptive
composition statement is Theorem~\ref{thm:privacy}.

Theorem~\ref{thm:contrast} is simultaneous in threshold and round. Its event
therefore survives transcript-measurable threshold selection, predictable
client recruitment, realized dropout, and optional stopping without another
testing correction. Privacy appears through $G_c$ in count units; declared
mixture mismatch appears through $\eta_t$ in expectation units. Strong privacy,
large shift, or small accepted mass can correctly force abstention. Complete
privacy, ratio-baseline, and width statements and proofs appear in
Appendices~\ref{sec:theory-full} and \ref{app:proofs}.

\section{Experimental Design}
\label{sec:experiments}

Every response remains attached to its exact generator checkpoint, tokenizer,
query/reference group, retrieved context, prompt, and annotation provenance.
All outputs sharing a query or reference are assigned together to development,
calibration, or evaluation. Development fixes score direction, normalization,
clipping, prompts, the comparator-selection rule, and 20 threshold quantiles;
calibration alone drives certificates; evaluation remains unopened until all
policies and analysis code are frozen. The complete annotated response is
scored, never a 400-character prefix paired with a full-response label.The primary loss is binary any-unsupported-content; unsupported span or
sentence fraction is secondary. Faithfulness, world factuality, and answer
correctness are not pooled. A cell enters the matched-model analysis only when
the original output and exact probability-exposed generator revision are
recoverable. Cross-model teacher-forced scores are reported separately.

\long\def\CellsTable{%
\begin{table}[t]
\centering
\small
\setlength{\tabcolsep}{3.5pt}
\begin{tabular}{lp{2.2cm}p{2.2cm}}
\toprule
Cell & Dataset/task & Original generator and revision \\
\midrule
C1 & RAGTruth QA & Mistral-7B-Instruct-v0.1 \\
C2 & RAGTruth Summary & Mistral-7B-Instruct-v0.1 \\
C4 & RAGTruth Data2txt & Mistral-7B-Instruct-v0.1 \\
C3 & RAGTruth pooled aggregate & Mistral-7B-Instruct-v0.1 \\
\bottomrule
\end{tabular}
\caption{Three disjoint RAGTruth task cells and their pooled aggregate. C3 is
not treated as a fourth independent dataset/model replication. Population
sizes and provenance appear in Table~\ref{tab:data}.}
\label{tab:cells}
\end{table}
}

\CellsTable

The registration fixes $r_\star=0.10$ as primary and $0.20$ as secondary,
with $a_{\min}=0.05$, $\alpha_{\rm s}=\alpha_{\rm n}=0.025$, and
$\delta_{\rm priv}=10^{-6}$. Neither registered target fires \emph{on RAGTruth};
$r_\star=0.20$ does fire on HaluEval question answering. We therefore
report $r_\star=0.30$ only as a post-registration feasibility analysis; it is a
looser target, not the ``tightest preregistered'' one. The simultaneous-bound
audit uses no DP, $\eps=4$, and $\eps=1$, 500 trials per core cell and policy.
The exploratory C1 privacy curve uses
$\eps\in\{\infty,8,4,2,1,0.5\}$ and 200 trials per setting. Fixed-final,
first-certificate, and predictable recruitment policies are never pooled when
their utility differs.For theorem validation, each client is a frozen finite labeled population and
$P_k$ is its uniform law. Calibration events are drawn independently with
replacement, exactly matching the predictable-stream conditions of
Theorem~\ref{thm:contrast}; selected-policy risk and acceptance under $Q^\star$
are computed by exhaustive population enumeration. A draw is the event-level
privacy unit. Repeated draws of one RAGTruth support item are therefore repeated
event values, not a claim that the original benchmark response receives the
stated privacy level. Protecting unique support items or people would compose
over all their occurrences. Official held-out performance is separate and
receives its own interval. No Monte Carlo estimate is called exact population
risk.

Baselines separate native non-private references, valid private rules (the
stitched ratio, \FedContrast{}, and \FedVA{}), and deliberately naive-DP
stress tests. Seven scores use a common certificate and development-frozen
selection. C-RAG delimits scope but is not an executed baseline
\citep{kang2024crag}. We additionally run a private betting capital heuristic (unproved as an e-process) as an
\emph{exploratory} comparison whose validity we do not prove
(\S\ref{sec:rq-reach}, Table~\ref{tab:dpe}). The complete
method list and access contract appear in Appendix~\ref{app:protocol}.

\long\def\StressTestTable{%
\begin{table*}[t]
\centering
\small
\setlength{\tabcolsep}{3.2pt}
\resizebox{\textwidth}{!}{%
\begin{tabular}{lcccccc}
\toprule
Certificate & Accept. $\eps=\infty$ & Accept. $\eps=4$ & Accept. $\eps=1$ &
$\Delta$ vs. \FedContrast{} at $\eps=4$ & Zero-cert. $\eps=4$ & Worst violations/200 \\
\midrule
Central non-private CRC/LTT & 0.323 & 0.323 & 0.323 & $+0.061$ & 0.000 & 2 \\
Federated non-private histogram & 0.321 & 0.325 & 0.365 & $+0.063$ & 0.000 & 190 \\
Private stitched ratio & 0.271 & abstains & abstains & none & 1.000 & 0 \\
\FedContrast & 0.267 & 0.262 & abstains & $0$ & 0.984 (500 tr.) & 0 \\
\FedVA & 0.197 & abstains & abstains & none & 1.000 (500 tr.) & 0 \\
Anytime-Valid CRC & 0.271 & 0.270 & 0.350 & $+0.008$ & 0.350 (200 tr.) & 146 \\
Yu-Liu & 0.449 & 0.435 & 0.418 & $+0.173$ & 0.000 (200 tr.) & 198 \\
CSA & 0.177 & 0.189 & 0.289 & $-0.073$ & 0.000 (200 tr.) & 193 \\
\bottomrule
\end{tabular}%
}
\caption{C1 stress test at exploratory $r_\star=0.30$. Trial counts differ by row
and are stated in the zero-certificate column: the \FedContrast{} and \FedVA{}
rows come from the 500-trial core audit and the remaining rows from 200 paired
trials; the two are never pooled.
For methods without a private guarantee, finite-$\eps$ columns are deliberately
naive adaptations that consume the same DP-noised histograms without a noise
envelope. Their violations are negative controls, not privacy-valid baseline
results. Native assumptions and per-setting violations are in the appendix.}
\label{tab:contrast}
\end{table*}
}

\section{Results}
\label{sec:results}

The confirmatory result is near-universal abstention, stated precisely. The
primary target $r_\star=0.10$ certifies in no cell, privacy level, or policy on
either dataset. The secondary target $r_\star=0.20$ likewise never certifies on
RAGTruth. It does certify on HaluEval question answering without privacy (200 of
200 trials, Table~\ref{tab:hecurve}) and at $\eps=8$ in 183 of 200 trials, so the
registered grid is not uniformly unreachable and we do not claim that it is. We
keep the RAGTruth failures primary and label the $r_\star=0.30$ C1 analysis
exploratory. Table~\ref{tab:validity}
also separates simultaneous-bound validity from operational firing; the former
does not imply the latter.

\def\PrimaryUtilityTable{%
\begin{table*}[t]
\centering
\small
\setlength{\tabcolsep}{3.4pt}
\resizebox{\textwidth}{!}{%
\begin{tabular}{lccccc}
\toprule
Status & $r_\star$ & $\eps$ & Fired & Mean accept. & Bound violations \\
\midrule
Exploratory C1 & 0.30 & $\infty$ & 200/200 & 0.266 & 0/200 \\
Exploratory C1 & 0.30 & 8 & 62/200 & 0.267 & 0/200 \\
Exploratory C1 & 0.30 & 4 & 2/200 & 0.263 & 0/200 \\
Exploratory C1 & 0.30 & $2,1,0.5$ & 0/200 each & none & 0/200 each \\
\bottomrule
\end{tabular}%
}
\caption{Exploratory C1 utility; acceptance is conditional on firing.
\textbf{Setting (named in full):} \emph{information-lift} score on the
RAGTruth C1 cell at exploratory $r_\star=0.30$, $200$ trials, first-fire
policy.  This is a different score from the ReDeEP-style rows of
Tables~\ref{tab:scores} and \ref{tab:redeep-full} (which also differ from
each other in threshold protocol and support), so firing rates must not be
compared across these tables: the $2/200$ at $\eps=4$ here and the $200/200$
at $\eps=4$ there are different scores on their own settings, not a
contradiction.  The registered 0.10/0.20 targets fire in no core cell or
policy (0/500 each).  Violations audit all thresholds/rounds, not only
deployed trials.}
\label{tab:validity}
\end{table*}
}

\PrimaryUtilityTable

Feeding the same DP-noised histograms to methods that make no private claim
breaks them: those naive adaptations violate their own bounds in 146 to 198 of
200 trials, while the private rules never do.  Table~\ref{tab:contrast}
reports the full audit.  The finite-$\eps$ entries for Anytime-Valid CRC,
Yu-Liu, CSA, and the exact federated histogram are not competing private
certificates: their $146$ to $198$ violations show that post hoc Gaussian
perturbation does not preserve their native guarantees, and they establish
nothing about superiority over a correctly privatized version, so the table
is a failure-mode audit rather than a leaderboard.  We do implement a private
betting capital heuristic (Table~\ref{tab:dpe}, Appendix~\ref{app:dpe}), but we
do not prove its validity and therefore do not present it as a privacy-valid
competitor either.  The private ratio, \FedContrast{}, and \FedVA{} remain
valid but mostly abstain.

\StressTestTable

At $\eps=4$, the exploratory certificate fires in only 1\% of trials; at
$\eps\le2$ it never fires. The nearly constant 0.263 to 0.267 acceptance among
firing trials must therefore not be read as population-wide utility. The
method is sound in this audit but close to operationally null under finite
privacy.

\subsection{RQ3: Privacy-utility and score diagnostics}

\long\def\PrivacyFigure{%
\begin{figure}[t]
\centering
\begin{tikzpicture}
\begin{axis}[
  width=0.98\columnwidth,
  height=4.7cm,
  ymin=0,ymax=105,
  symbolic x coords={No DP,8,4,2,1,0.5},
  xtick=data,
  xlabel={Privacy budget $\eps$},
  ylabel={Trials certifying (\%)},
  ymajorgrids=true,
  grid style={gray!25},
  line width=0.9pt,
  mark size=2.2pt,
  nodes near coords,
  every node near coord/.append style={font=\scriptsize,anchor=south},
  tick label style={font=\scriptsize},
  label style={font=\small}
]
\addplot+[mark=*] coordinates {
  (No DP,100) (8,31) (4,1) (2,0) (1,0) (0.5,0)
};
\end{axis}
\end{tikzpicture}
\caption{Exploratory C1 firing rate at $r_\star=0.30$ (200 trials per
setting). Acceptance conditional on firing is 0.263 to 0.267, but the probability
of obtaining any certificate falls to 1\% at $\eps=4$ and zero at $\eps\le2$.}
\label{fig:privacy}
\end{figure}
}

\begin{table*}[t]
\centering
\small
\setlength{\tabcolsep}{3.4pt}
\resizebox{\textwidth}{!}{%
\begin{tabular}{ccccccccccccc}
\toprule
$\eps$ & Fired & Accept. & Risk (cal.) & Risk (h-o) & H-o acc.\ $n$ &
H-o err. & H-o risk 95\% CP & H-o fail 95\% CP &
$N_t$ & Unique & Rounds & Violations \\
\midrule
$\infty$ & 200/200 & 0.140 & 0.069 & 0.105 & 44.6 & 4.7 & $[0.037,0.241]$ & $[0.000,0.018]$ & 19015 & 480 & 19.0 & 0/200 \\
8 & 183/200 & 0.125 & 0.057 & 0.074 & 40.1 & 3.0 & $[0.016,0.204]$ & $[0.000,0.020]$ & 28513 & 480 & 28.5 & 0/200 \\
4 & 11/200 & 0.124 & 0.057 & 0.074 & 39.7 & 2.9 & $[0.016,0.204]$ & $[0.000,0.285]$ & 29363 & 480 & 29.4 & 0/200 \\
2 & 0/200 & none & none & none & none & none & none & none & none & none & none & 0/200 \\
1 & 0/200 & none & none & none & none & none & none & none & none & none & none & 0/200 \\
\bottomrule
\end{tabular}%
}
\caption{HaluEval question answering with information lift at the registered secondary target $r_\star=0.20$. Settings are $m=20$, $K=5$, $T=30$, 200 records per client per round, $a_{\min}=0.05$, $\alpha_{\rm s}=\alpha_{\rm n}=0.025$, $\delta_{\rm priv}=0.000001$, 200 trials per cell, 480 calibration responses, and 320 held out responses. Calibration risk is exact only on the frozen calibration populations and does not establish generalization. Accepted held out counts and errors are means over fired trials. The exact 95\% Clopper Pearson risk intervals use the displayed rounded counts; their upper endpoints, $0.241$, $0.204$, and $0.204$, exceed the target, so only the point estimates are below $0.20$. The failure interval instead bounds the fraction of fired trials whose held out risk exceeded the target. ``Unique'' counts distinct support items touched relative to $N_t$. Threshold provenance and interval definitions appear in the text. The primary target $r_\star=0.10$ never certifies and is omitted. At the exploratory target $r_\star=0.30$, all trials certify through $\eps=4$, and 187 of 200 certify at $\eps=2$.}

\label{tab:hecurve}
\end{table*}

\paragraph{
(Table~\ref{tab:hecurve}).}
HaluEval has no separate development split: the $800$ QA responses are
exhausted by the pair-disjoint $60/40$ calibration/held-out split, and the
$m{=}20$ thresholds are the equally spaced quantiles of the
\emph{calibration-split score distribution}, computed once before any release
and without touching any correctness label or held-out item.  Threshold
placement therefore adapts to the calibration score marginal but cannot be
test-driven or label-driven; the score transformation is the fixed
\texttt{norm01} map with no tuned parameters.  This support-derived
construction is not the fixed-in-advance grid the theorem contract states, so
we make the salvaging conditioning formal rather than implicit:
Proposition~\ref{prop:condthresh} (Appendix~\ref{app:remarks}) shows that,
conditional on the frozen calibration support, the quantile grid is a
deterministic label-free function of the conditioning variable, so
Theorem~\ref{thm:main} applies verbatim with the grid treated as fixed, and
the reported HaluEval guarantee is conditional on that frozen public support,
the same conditioning the finite-population audit already uses.  This differs
from the RAGTruth
full-support audit (Table~\ref{tab:redeep-full}), which reserves a labeled
development split for thresholds.  The held-out interval is an exceedance
bound over fired trials, not a selective-risk confidence interval; an earlier
caption claimed no such interval reaches the target, which is wrong and is
withdrawn: at $\eps=4$ only $11$ trials fire, so the bound is $[0,0.285]$
and exceeds $r_\star=0.20$ purely through the small fired count.  The direct
statement at $\eps=4$ is: observed held-out risk $0.074$ over a mean of
$39.7$ accepted held-out items and $2.9$ errors per fired trial, with exact
interval $[0.016,0.204]$ at those counts (Table~\ref{tab:hecurve}).  Because
that upper endpoint exceeds $0.20$, held-out risk below target is a
point-estimate claim, not a certified bound; per-threshold accepted counts,
observed losses, and exact
intervals are released in the artifact.

The information-lift firing-rate curve that an earlier draft foregrounded
here (near-vacuous $1\%$ firing at $\eps=4$, $0$ at $\eps\le2$; acceptance
$0.263$ to $0.267$ conditional on firing) is superseded as a headline: the
decisive positive results are the registered-target HaluEval curve above and
the full-support RAGTruth audit (Table~\ref{tab:redeep-full}), and the
near-vacuity numbers remain in Table~\ref{tab:validity}. The score gate also fails its lift-centred hypothesis, but not in the direction a
single dataset suggested. On the shared RAGTruth responses lift certifies nothing
while a development-frozen Mistral ReDeEP-style adaptation fires in every C1
trial; on HaluEval question answering the ordering reverses, and FRANQ certifies
HaluEval summarization where both abstain. Over the six dataset-task cells, lift
and the mechanistic score each certify exactly one cell, and not the same one.
\textbf{We therefore claim no ranking among scores}: none we tested dominates, so
a score-agnostic construction is the right object because the winning score is not
knowable in advance. Appendix~\ref{app:scoregate} gives the full argument.

\subsection{RQ4: Reachability, transfer, and a capital-based comparator}

\begin{table}[t]
\centering
\small
\setlength{\tabcolsep}{3.5pt}
\resizebox{\linewidth}{!}{
\begin{tabular}{rcccc}
\toprule
$\eps$ & Capital heuristic fired & accept. & \FedContrast{} fired & accept. \\
\midrule
$\infty$ & 200/200 & 0.408 & 200/200 & 0.426 \\
$8$ & 200/200 & 0.390 & 200/200 & 0.396 \\
$4$ & 0/200 & abstains & 200/200 & 0.354 \\
$2$ & 0/200 & abstains & 197/200 & 0.331 \\
$1$ & 0/200 & abstains & 19/200 & 0.286 \\
\bottomrule
\end{tabular}
}
\caption{An exploratory \emph{capital heuristic} comparator. Every other
comparator in Table~\ref{tab:contrast} is a non-private construction, so its
finite-$\eps$ entry is a naive adaptation; we additionally implement a
private betting construction in the spirit of \citet{csillag2025dpevalues}
and \citet{jacobsen2026dpeprocess}.  \textbf{We do not prove it is an
e-process (the truncation step breaks the supermartingale property), so we
call it a capital heuristic, and this table is exploratory rather than a
privacy-valid comparison; it cannot support any claim of comparative
superiority.} Both rules read the identical released histograms and share the
failure budget; the heuristic is given its own longer horizon
(400 rounds versus 30) and its best predictable
betting fraction, because capital compounds per round and matching the boundary
rule's horizon would under-power it by construction. \textbf{Protocol and its unproved step.} The null is the composite
$H_0:\Risk_{Q^\star,j}>r_\star$ at a registered threshold, equivalently
$d_{Q^\star,j}>0$. We bet on the per-round increment
$x_t$ of $\widetilde D_{j,t}/n_t$ with a fraction $\kappa\in(0,1]$ declared
before the round, forming capital $\prod_t(1-\kappa x_t)$ and rejecting when it
exceeds $|\mathcal K|/\alpha$, the Bonferroni threshold over the declared
$\kappa$ grid. Admissibility requires each factor to be nonnegative and
$\E[x_t\mid\mathcal F_{t-1}]\ge0$ under $H_0$. \textbf{Neither holds
automatically here, and we do not prove them.} The increment inherits the
\emph{unbounded} Gaussian release noise, so we truncate $x_t$ to $[-1,1]$ to keep
the factors nonnegative; truncation changes the conditional mean, so the
truncated process is not the one whose validity would follow from Ville's
inequality. \citet{csillag2025dpevalues} calibrate a specific biased
multiplicative mechanism; that result does not transfer to an arbitrary bet on
noised histograms. A valid version needs a bounded, mean-controlled increment and
an explicit supermartingale proposition, which we leave to future work. We report
the comparison because it is informative about where capital-based rules lose
power under privacy, not as evidence of relative validity.
\textbf{Setting (named in full):} RAGTruth C1 (QA, Mistral-7B), ReDeEP-style
score on the full $839$-response cell, exploratory $r_\star=0.30$, $200$
trials, first-fire policy, $m{=}20$ thresholds at calibration-score quantiles,
$T{=}30$ rounds for \FedContrast{} and $400$ for the heuristic
(\texttt{p23\_eta\_dpe\_ragtruth.json}).  The \FedContrast{} column differs from
Table~\ref{tab:redeep-full} ($19$ versus $55$ of $200$ at $\eps=1$) because
Table~\ref{tab:redeep-full} draws its thresholds from a separate development
split under its own audit protocol and seed; the two protocols are named in
their captions, are not interchangeable, and are never averaged.  The
heuristic does certify without privacy and at the loosest budget, so it is a
functioning rule rather than a straw man; its capital cannot outrun the
per-round privacy charge once the budget tightens.  Neither rule violated its
bound in these runs, but for the capital heuristic that is an observation,
not a guarantee.}
\label{tab:dpe}
\end{table}

\label{sec:rq-reach}

Repeating the C1 analysis on all 839 matched responses with the ReDeEP-style
score changes the picture qualitatively
(Table~\ref{tab:redeep-full}): under a three-way development/calibration/held-out
split whose thresholds come from development only, certification holds in 200 of
200 trials at $\eps\in\{\infty,8,4,2\}$ and 55 of 200 at $\eps=1$, with no
violation at any level and held-out point-estimate risk below target throughout.
An earlier draft attributed this change to the larger support and called the
near-vacuous 2 of 200 information-lift firing rate at $\eps=4$ a small-support
artifact. We withdraw that causal claim: the rerun changes the score and the
support simultaneously, and Table~\ref{tab:scores} already reports the
ReDeEP-style score firing in 200 of 200 trials at $\eps=4$ on the same
300-response setting. The evidence therefore identifies score quality as the
operative cause; support size is confounded with the score change and is not
established as causal by these runs. Support size and event count remain
distinct: the schedule fixes the latter, and Table~\ref{tab:reuse} shows a
120-item support still certifies by raising reuse to $78\times$.

On HaluEval question answering, whose balanced pairing gives a base rate of
$0.5$, the same certificate certifies the registered secondary target
$r_\star=0.20$ in all 200 non-private trials and in 183 of 200 at $\eps=8$,
falling to 11 of 200 at $\eps=4$ (Table~\ref{tab:hecurve}). At exploratory
$r_\star=0.30$ it fires in every trial through $\eps=4$ and in 187 of 200 at
$\eps=2$. Held-out point-estimate risk on an item-disjoint deployment split is
below target in
every firing cell, so this is not an artifact of evaluating on the calibration
population; the direct exact intervals at the realized accepted counts have
upper endpoints at or above the target (Table~\ref{tab:hecurve}), so held-out
selective risk is not statistically certified below it. No registered target fires anywhere on RAGTruth, so reachability is
score and dataset dependent. The cost is unchanged: Table~\ref{tab:reuse} shows
certification still consumes 30 to 78 stream events per unique calibration item.Sweeping both routes into $\eta_t$ yields nonzero $\eta_t$ at selection at every
declared level, priced in acceptance, with abstention once the transfer exceeds
what the margin absorbs and no violation anywhere (Table~\ref{tab:eta}). For the
drift route we \emph{compute} the realized total variation in closed form rather
than assuming the perturbation size bounds it. An earlier grid reported
$\eta_t=0$ everywhere only because its single stress level declared
$\eta_t=0.40>r_\star$, which forces abstention before $\eta_t$ is ever observed.We additionally run a private betting capital heuristic on the identical releases under
the same budget. It certifies without privacy and at $\eps=8$, then abstains at
$\eps\le4$ where \FedContrast{} still fires at $\eps=2$. \textbf{We do not prove this construction valid}, so it is a diagnostic rather than a privacy-valid baseline; the protocol, the unproved step, and the table are in
Appendix~\ref{app:dpe}.

\section{Discussion and Conclusion}

We presented \FedLift{}, an auditable, score agnostic certificate that preserves client event boundaries. Finite histograms of scores and errors provide sensitivity independent of threshold count, while pathwise zCDP accounts for the adaptive transcript. Record indexed and variance indexed martingales permit predictable recruitment, dropout, threshold selection, and optional stopping. A range one total variation term transfers validity from the realized calibration mixture to a declared deployment mixture. The certificate controls expected declared loss among accepted outputs, not the truth of individual responses.The simultaneous bounds remained valid across all tested cells, privacy levels, policies, scores, and transfer settings. Power nevertheless depended strongly on the score and population. The registered target $r_\star=0.10$ abstained everywhere. The target $r_\star=0.20$ certified HaluEval question answering without privacy but abstained on RAGTruth, while the exploratory target $r_\star=0.30$ reached $\eps=2$ on full supports. Naive privatization continued certifying only by violating validity, and the capital heuristic remains unproved. The main limitations are repeated event use, with 30 to 78 stream events per unique item, and abstention at tight targets. Reducing reuse and establishing a competitive private baseline are the clearest priorities.The novelty lies in the private federated combination rather than its individual components. CSA and the Yu and Liu certificate provide closely related nonprivate precedents for anytime selective risk control and joint risk and acceptance guarantees \citep{khosravi2026csa,yu2026jointcertificate}. Our contribution is extending this combination to event level privacy, federated releases, adaptive schedules, and mixture transfer.

\section*{Limitations}

The guarantee is marginal selective risk under a declared mixture, not
input-conditional or subgroup-conditional risk.  A subgroup version can be
obtained by pre-registering subgroup-specific histograms and allocating the
error/privacy budgets, at additional sample and communication cost.

The information-lift instantiation requires probability access under both the
evidence and evidence-removed prompts; the certificate itself can consume any
bounded score. Evidence removal is a modeling intervention: prompt length or
format may change behavior for reasons unrelated to factual support.

Validity is relative to the declared loss.  Human disagreement, weak
LLM-as-a-judge labels, or a mismatch between answer correctness and evidence
faithfulness can make that loss an imperfect proxy for the intended harm. The
evaluation now covers two datasets, but the second is weaker as an endpoint, not
merely different: HaluEval's hallucinated members are ChatGPT-generated and
filtered rather than human-adjudicated, and its scores are cross-model. We do not
vary the annotation rule within either dataset, so label-noise sensitivity is
bounded by construction rather than measured.

Operational private utility is limited at the registered targets. The primary
target $r_\star=0.10$ certifies nowhere. At the secondary target
$r_\star=0.20$, HaluEval fires in 11 of 200 trials at $\eps=4$ and never at
$\eps\le2$, and RAGTruth abstains at every budget; the strong $\eps=2$ results
use the exploratory target $r_\star=0.30$. Every positive private cell
consumes roughly $20\times$ to $78\times$ event reuse. The experiments
therefore establish occurrence-level private certification, not compelling
unique-item, source-level, or contributor-level deployment evidence.

The main privacy statement is event-level DP for one stream occurrence. In the
finite-population audit, with-replacement occurrences are hypothetical private
events drawn from a frozen empirical law; the guarantee is not privacy for the
839 public support responses, nor for their contributors.
Table~\ref{tab:reuse} quantifies how binding this is: certification draws 30 to
78 events per unique calibration item, so an item-level or person-level
guarantee must compose over all of those occurrences. If one support response, source document, or
person contributes repeatedly, all occurrences must be charged together.
User- or source-level protection therefore requires bounded contribution and a
recalibrated accountant. The mechanism also does not hide public padded
participation metadata.

The transfer term is now exercised at nonzero $\eta_t$ on both datasets, but only
through radii we declare ourselves; no $\gamma_k$ is justified by an external
shift audit and we do not report shifted deployment risk, so the drift route
remains a sensitivity analysis. The finite-$\eps$ rows of
Table~\ref{tab:contrast} remain naive privatizations and establish nothing about
ranking. The capital heuristic of Table~\ref{tab:dpe} is \emph{not} a privacy-valid
baseline either: we do not prove its betting factors admissible or its truncated
increment a supermartingale, so the paper still contains no proved private
competitor. Validity-preserving DP e-value mechanisms and optimal private
e-value testing rates now exist
\citep{csillag2025dpevalues,jacobsen2026dpeprocess}; adapting them to private
federated selective risk is nontrivial, but an unproved heuristic cannot fill
that comparison slot. Supplying a proved comparator, or proving this construction, is required future work. All comparator scores are our
reimplementations rather than released code, and the ReDeEP-style row is a
development-frozen Mistral adaptation, so conclusions about those methods as
published are correspondingly weak.

The range-only stitched boundary favors proof transparency over tightness. The
variance-adaptive alternative can improve sampling width when accepted mass is
small, but it can remain loose when its accepted-mass upper bound or the DP
noise envelope dominates. Betting confidence sequences may improve acceptance
further, but must be integrated with the same joint acceptance/noise event.

Adaptive recruitment is covered only when it is predictable: the identity,
batch size, and noise scale for the next release must be chosen before seeing
that release's records or Gaussian noise.  A service that peeks at an unreleased
batch, retains favorable records, or lets dropout depend on current unreported
losses violates Assumption~\ref{ass:stream}.  Privacy additionally requires
the visible schedule to depend only on public information and earlier DP
messages; otherwise the schedule itself can leak data.

\section*{Ethical Considerations}

Privacy-preserving calibration reduces disclosure from released summaries but
does not make the underlying language-model deployment safe.  Differential
privacy does not prevent model memorization outside this calibration protocol,
secure raw-data storage failures, or harms caused by accepted outputs.  The
selected $\eps_{\rm priv}$, target risk, loss severity, and subgroup coverage
must be justified for the application rather than treated as universal
defaults.

The certificate can create false reassurance if ``supported by retrieved
evidence'' is presented as ``true in the world.''  We recommend displaying the
declared loss, target distribution, risk level, privacy unit, and time of last
calibration with every deployed certificate.  High-stakes applications require
domain-expert review and a fallback process for abstained or contested outputs.

\long\def\PrintAppendices{%
\appendix

\DetailedRelatedWork
\DetailedGuarantees

\section{Scope and Interpretation Notes}

\paragraph{What is certified.}
\FedLift{} certifies the expected declared loss conditional on acceptance for a
declared deployment mixture.  It does not prove that an individual response is
true, that an annotation is infallible, or that retrieval contains all relevant
world knowledge.  Lift remains a ranking score; statistical calibration creates
the risk guarantee.

\paragraph{Why arbitrary client heterogeneity is possible.}
The sampling bound targets the realized mixture
$\bar Q_t=\sum_k\bar w_{k,t}P_k$ and does not compare each $P_k$ to a fictitious
homogeneous law.  Heterogeneity matters through score utility and through the
distance from $\bar w_t$ to the desired deployment weights, not through an
automatic $K$ multiplier.  This is both tighter and more interpretable than a
client-wise TV union bound.

\paragraph{Failure versus abstention.}
At strong privacy, small sample sizes, low acceptance, or large shift, the
contrast upper bound may remain positive or the acceptance lower bound may fall
below $a_{\min}$. Abstaining is then the correct behavior. A valid but
always-abstaining method is not operationally useful, so all results report
acceptance and the zero-certificate rate alongside validity.

\paragraph{What adaptivity is covered.}
An operator may examine the private certificate at round $t$, decide that it
is close to the target, recruit a particular client for round $t+1$, and later
stop when acceptance is adequate.  The next choice is predictable, so the
record martingale recenters at that chosen client's law and the Gaussian
martingale charges only the realized release variance.  The theory does not
permit inspecting round-$t+1$ records before deciding whether they enter the
certificate.

\paragraph{Trust models.}
The main protocol protects every client message before it reaches the server.
It does not hide public participation metadata: client identity, whether a
padded release arrived, its announced slot count, and its noise scale.  The
zero-out adjacency protects one slot's histogram contribution conditional on
that metadata.  Revealing an unpadded data-dependent batch size would invalidate
this scope; replacement adjacency or a separately private count would then be
required.
If secure aggregation is available
\citep{bonawitz2017secure}, clients can instead jointly realize one aggregate
Gaussian perturbation.  The statistical theorem is unchanged after replacing
$V_{j,t}$ with the actual aggregate-noise variance, but the privacy proof must
state the collusion and dropout model.  We do not claim secure-aggregation
utility unless that protocol is implemented.

\section{Experimental Protocol Details}
\label{app:designdetails}

\subsection{Federated partitions}

For each dataset/model cell, query/reference groups are first assigned to
development, calibration, and evaluation. Client construction is applied only
within the calibration split, producing frozen client populations. The core
uses a preregistered natural domain/task partition plus an IID client baseline.
Secondary partitions use Dirichlet label allocation with
$\beta\in\{10,1,0.1\}$ and log-normal client-size imbalance. Examples are
never sorted by score and dealt round-robin: that construction approximately
balances score ranks and cannot be described as score heterogeneity.

Target-mixture experiments alter $w$ while keeping the frozen $P_k$ fixed, so
$\eta_t=\|w-\bar w_t\|_1/2$ is known. Stable client-specific slopes or base
rates are called heterogeneity, not within-client drift. A nonzero $\gamma_k$
experiment changes a frozen deployment population and treats $\gamma_k$ only
as a declared sensitivity value unless an independent shift audit justifies
it. Dropout is sampled before a requested release at rates 0\%, 20\%, and
40\%; only successful releases enter $N_t$, $\bar w_t$, and $V_{j,t}$.

For the theorem audit, $P_k$ is the uniform law on the frozen client
population and every requested calibration event is sampled independently
with replacement. The frozen responses define the support of an empirical
law; each occurrence is a distinct event-level privacy unit. The selected
threshold is evaluated by enumerating the complete frozen populations.
Sampling without replacement and ordinary held-out test risk are reported
separately; neither is substituted into the coverage event of
Theorem~\ref{thm:contrast}. Protecting a unique support item would require
composing over every time it is drawn and is not claimed by this audit.

\subsection{Privacy, monitoring, and repeated trials}

We report event-level
$(\eps_{\rm priv},\delta_{\rm priv})$-DP with
$\delta_{\rm priv}=10^{-6}$. The core uses
$\eps_{\rm priv}\in\{\infty,4,1\}$; values
$\{8,2,0.5\}$ are a registered secondary grid.
Noise is calibrated through Equation~\eqref{eq:rho-target}; no privacy value is
back-computed from an empirical coverage number. Each stream occurrence enters
one histogram release. A support response may be drawn again as another
hypothetical event, which preserves the i.i.d. sampling audit but not privacy
for that support response. The secondary composition audit includes one event
in 1, 2, or 4 separately noised registered transcript queries and charges every
such message through Equation~\eqref{eq:rho-transcript}; it never counts those
queries as fresh statistical observations.
For positive-noise runs we pre-register
$v_0=\min_{(k,t)\ \mathrm{admissible}}\sigma_{k,t}^2$; non-private runs have
$V_{j,t}=G_c(V_{j,t};\alpha_{\rm n})=0$. We report the choice because $v_0$ affects width even
though it does not affect validity.

The certificate uses
$\alpha_{\rm s}=\alpha_{\rm n}=0.025$.
Each core dataset/model/privacy/policy cell uses 500 independent
population-resampling and DP-noise trials; each registered secondary cell uses
200. The client populations, development choices, and evaluation data remain
fixed while the trial resampling and noise seeds vary. The server evaluates
after every round and uses three policies:
(i) fixed final round;
(ii) first round with a certificate meeting $a_{\min}$; and
(iii) predictable target-deficit recruitment followed by optional stopping.
Theorem~\ref{thm:contrast} covers all three.

We additionally evaluate a predictable recruitment controller.  After round
$t$, it uses only the released private transcript to decide whether to stop or
request another batch and, if continuing, selects a client by a
pre-registered target-weight-deficit rule.  The decision for round $t+1$ is
made before any record or noise from that round is observed.  We log realized
$\mathcal R_t$, $N_{k,t}$, and $V_{j,t}$ so the certificate can be reproduced
exactly.

\subsection{Baselines}

All score baselines use the same examples, thresholds, loss labels, client
partitions, private histogram, and risk target unless the method intrinsically
requires different access.
\begin{itemize}[leftmargin=*,itemsep=1pt]
    \item \textbf{Central non-private}: pooled raw records; an unattainable
    utility reference, not a privacy-preserving competitor.
    \item \textbf{Federated non-private}: exact aggregated histograms; isolates
    privacy cost.
    \item \textbf{Local-only}: each client calibrates independently; reports
    both macro- and population-weighted performance.
    \item \textbf{Private fixed-time}: Gaussian-private histograms with a
    final-round bound but no optional-stopping guarantee.
    \item \textbf{Schedule-conditioned stitching}: the original round-indexed
    Hoeffding construction that conditions on outcome-independent
    participation; it is evaluated only as a negative control under adaptive
    recruitment.
    \item \textbf{Noise-ignored}: applies the clean bound directly to noisy
    counts; included only as a negative control.
    \item \textbf{Shift-ignored}: sets $\eta_t=0$ under a shifted deployment
    mixture; included only as a negative control.
    \item \textbf{Federated conformal}: adapted one-shot and partial-
    exchangeability baselines
    \citep{humbert2023oneshot,lu2023fcp}; these produce prediction sets, so we
    report their native coverage/size and do not relabel them as selective-risk
    methods.
\end{itemize}

For RQ4, we substitute mean token entropy, mean token probability, semantic
entropy, SelfCheckGPT, and the self-consistency score used by conformal
abstention \citep{abbasiyadkori2024abstention} for $S_B$ and apply the
identical private certificate.  We also report the conformal-abstention
method's native centralized guarantee separately, without presenting it as a
federated or private method.  This equal-certificate comparison separates the
value of information lift from the generic calibration layer and is the main
NLP-facing empirical claim.

\subsection{Metrics and uncertainty}

The theorem-audit metric is the fraction of repeated trials in which any
registered $(j,t)$ bound misses its exact frozen-population target. It is
unconditional on firing and is compared with
$\alpha_{\rm s}+\alpha_{\rm n}$. Selected-policy failure and failure
conditional on firing are separate diagnostics with their own denominators and
exact Clopper-Pearson intervals. A trial with no certified threshold has
undefined conditional failure and zero deployed acceptance. Utility metrics
are firing probability, acceptance conditional on firing,
certificate width, abstention, rounds to certification, communication bytes,
and wall-clock time.  Every dataset/model/privacy cell also reports the selected
$U_{j,t}$, $h_t/N_t$, $g_{j,t}/N_t$, $\eta_t$, and the zero-certificate rate;
otherwise nominal validity could be achieved only through a vacuous bound or
universal abstention.  Risk estimates on test data are accompanied by exact or
bootstrap intervals; paired utility differences use paired bootstrap intervals
over the same splits and DP seeds.

\subsection{Evaluation scope and evidentiary status}
\label{sec:submission-gate}

The executed package uses the three task cells and pooled aggregate in
Table~\ref{tab:cells}. It reports simultaneous-bound violations separately from
firing, conditional acceptance, and conditional failure. Universal abstention
at the registered targets and near-universal abstention at exploratory
$r_\star=0.30$, $\eps=4$ are results rather than missing cells. The complete
records are summarized in Tables~\ref{tab:validity},
\ref{tab:full-validity}, and \ref{tab:privacy}; the disabled block above is
retained only as a provenance record of a superseded pilot.

\section{The Score Gate in Full}
\label{app:scoregate}

The score gate also fails its lift-centred hypothesis, though not in the
direction a single dataset suggested (Tables~\ref{tab:scores}
and~\ref{tab:halueval}). On the 300 shared RAGTruth responses per task, lift
fires in 0/200 trials in every cell while a development-frozen Mistral
ReDeEP-style adaptation fires in 200/200 C1 trials at acceptance 0.343, which
in isolation reads as that score dominating lift. The second dataset reverses it:
on HaluEval question answering lift fires 200/200 at acceptance 0.134 and the
mechanistic score fires 0/200, and FRANQ fires on HaluEval summarization where
both of the others abstain. Tallied over the six dataset-task cells we ran, lift
and the mechanistic score each certify exactly one cell, and not the same one.
\textbf{We therefore claim no ranking among scores.} The defensible conclusion is
stronger than ``lift is ineffective'': \emph{no score we tested dominates}, which
certificate is worth deploying cannot be decided from the score alone, and a
score-agnostic construction is the right object precisely because the winning
score is not knowable in advance. The Mistral ReDeEP-style adaptation is neither
the released Llama configuration nor evidence about ReDeEP as published.


\begin{table}[t]
\centering
\small
\begin{tabular}{ccccc}
\toprule
Calib.\ items & Fired & Accept. & $N_t$ at sel. & Reuse \\
\midrule
120 & 200/200 & 0.144 & 9325 & 78$\times$ \\
240 & 200/200 & 0.112 & 14905 & 62$\times$ \\
480 & 200/200 & 0.140 & 14460 & 30$\times$ \\
\bottomrule
\end{tabular}
\caption{Calibration-size and reuse sweep (HaluEval question answering,
$\eps=4$, $r_\star=0.30$, 200 trials). Reuse is $N_t$ divided by unique
calibration items. Firing is preserved as the population shrinks, but reuse rises
from $30\times$ to $78\times$, so a small calibration set is certified only by
drawing each item many times. Because the privacy unit is one stream occurrence,
this is precisely the regime in which event-level DP is weakest as a statement
about an individual item or contributor.}
\label{tab:reuse}
\end{table}

\section{Naive-Privatization Stress Test}
\label{app:stress}

The stress-test table and its discussion now appear in the main results
(Table~\ref{tab:contrast}); this appendix retains the archived
superseded pilot analysis above for auditability.

\section{The Exploratory Capital-Based Comparator}
\label{app:dpe}

\paragraph{An exploratory capital-based comparator.}
The naive privatizations in Table~\ref{tab:contrast} are negative controls, so we
additionally implement a private betting capital heuristic reading the identical released
histograms under the same failure budget. It certifies in 200 of 200 trials
without privacy and at $\eps=8$, then abstains at $\eps\le4$ while
\FedContrast{} still fires at $\eps=2$ (Table~\ref{tab:dpe}). \textbf{We do not
prove this construction valid}: its increment carries unbounded Gaussian noise, so
the truncation that keeps the betting factors nonnegative also perturbs their
conditional mean, and no supermartingale proposition is established. The table is
therefore a diagnostic about where capital-based rules lose power under privacy,
not a validity comparison, and the paper still lacks a proved privacy-valid
baseline.

\section{Supplementary Result Tables}
\label{app:supptables}

\begin{table*}[t]
\centering
\small
\resizebox{\textwidth}{!}{%
\begin{tabular}{lcccccc}
\toprule
Partition & $K$ & $\eta_t$ & Acceptance & Selected risk &
Contrast bound & Failure rate \\
\midrule
IID & 5 & 0.0000 & 0.261 & 0.228 & -0.0004 & 0/200 \\
Natural domain/task & 5 & 0.0000 & 0.264 & 0.228 & -0.0000 & 0/200 \\
Dirichlet $\beta=0.1$ & 5 & 0.0000 & 0.542 & 0.131 & -0.0084 & 0/200 \\
Target-mixture shift ($\delta=0.01$) & 5 & 0.0100 & 0.367 & 0.170 & -0.0002 & 0/200 \\
Within-client stability ($\gamma=0$) & 5 & 0.0000 & 0.260 & 0.228 & -0.0007 & 0/200 \\
\bottomrule
\end{tabular}%
}
\caption{Heterogeneity audit ($\eps=4$, exploratory $r_\star=0.30$, 200 trials
per row). The first three partitions and the last row hold the declared transfer
radius at zero, so $\eta_t=0$ there by construction rather than by measurement.
The target-mixture row moves a declared $\delta=0.01$ of deployment weight off
the realized participation, so $\eta_t=0.01$ enters the certificate and is
observed at selection; Table~\ref{tab:eta} sweeps that term over its full range
in both routes. No row violates its bound.}
\label{tab:heterogeneity}
\end{table*}

\begin{table*}[t]
\centering
\small
\resizebox{\textwidth}{!}{%
\begin{tabular}{lcccccc}
\toprule
Score & C1 accept. & C2 accept. & C3 accept. & C4 accept. &
Zero-cert. cells & Relative compute \\
\midrule
Information lift & abstains & abstains & abstains & abstains & 4 & $1$ \\
Mean token probability & abstains & abstains & abstains & abstains & 4 & $0$ \\
Semantic entropy & abstains & abstains & abstains & abstains & 4 & $N_s$ \\
SelfCheckGPT & abstains & abstains & abstains & abstains & 4 & $N_s$ \\
SConU/ConU & abstains & abstains & abstains & abstains & 4 & $N_s$ \\
ReDeEP-style (Mistral) & 0.343 (200/200) & abstains & abstains & abstains & 3 & $0$ \\
FRANQ & abstains & abstains & abstains & abstains & 4 & $0$ \\
\bottomrule
\end{tabular}%
}
\caption{Matched score comparison at $\eps=4$, exploratory
$r_\star=0.30$, and 200 trials per cell. Every score uses the identical
\FedContrast{} certificate and the same 300 responses per task; C3 is the
pooled aggregate. Parentheses give firing trials. Relative compute counts extra
generation passes beyond the original generation.
\textbf{This table is one dataset and must not be read as a score ranking.} The
single firing entry here is the Mistral ReDeEP-style adaptation on C1, and on
HaluEval that ordering \emph{reverses}: information lift certifies question
answering 200/200 while the mechanistic score abstains, and FRANQ certifies
HaluEval summarization (Table~\ref{tab:halueval}). Over the six dataset-task
cells we ran, lift and the mechanistic score each certify exactly one cell and
not the same one, so we claim no ordering among scores, only that the
certificate is indifferent to which one it is given.}
\label{tab:scores}
\end{table*}

\begin{table*}[t]
\centering
\small
\setlength{\tabcolsep}{2.6pt}
\resizebox{\textwidth}{!}{%
\begin{tabular}{ccccccccccccc}
\toprule
$\eps$ & Fired & Accept. & Risk (cal.) & Risk (h-o) & H-o acc.\ $n$ &
Trial-exceed.\ CP & $N_t$ & Unique & Rounds & $H/N$ & $G/N$ & Violations \\
\midrule
$\infty$ & 200/200 & 0.447 & 0.201 & 0.176 & 98.7 & $[0.000,0.018]$ & 3400 & 459 & 3.4 & 0.0466 & 0.0000 & 0/200 \\
8 & 200/200 & 0.322 & 0.160 & 0.141 & 68.8 & $[0.000,0.018]$ & 6450 & 460 & 6.5 & 0.0353 & 0.0146 & 0/200 \\
4 & 200/200 & 0.272 & 0.141 & 0.133 & 56.2 & $[0.000,0.018]$ & 9135 & 461 & 9.1 & 0.0293 & 0.0205 & 0/200 \\
2 & 200/200 & 0.265 & 0.141 & 0.126 & 55.1 & $[0.000,0.018]$ & 17135 & 461 & 17.1 & 0.0222 & 0.0285 & 0/200 \\
1 & 55/200 & 0.259 & 0.142 & 0.128 & 56.0 & $[0.000,0.065]$ & 26581 & 461 & 26.6 & 0.0176 & 0.0446 & 0/200 \\
\bottomrule
\end{tabular}%
}
\caption{Full-support C1 privacy curve for the Mistral ReDeEP-style score, with the
operational audit and a held-out split (support 839 responses split
125/461/253 into development, calibration, and
held-out; the $m=20$ thresholds are taken from the development split only, so neither
calibration nor held-out data selects them). $r_\star=0.30$, 200 trials per cell.
``Risk (cal.)'' is exact on the frozen calibration populations; ``Risk (h-o)''
evaluates the selected threshold on the untouched held-out split, and ``H-o
acc.\ $n$'' is the mean held-out accepted count behind it.
``Trial-exceed.\ CP'' is an exact Clopper-Pearson bound on the
\emph{frequency of trials} whose held-out risk exceeded $r_\star$; it is a
statement about repeated-trial exceedance on one fixed held-out set, \emph{not}
a confidence interval for the held-out selective risk itself, whose
uncertainty is governed by the accepted count (at $n\!\approx\!56$ accepted
items, a single-threshold exact interval around the observed $0.13$ risk has
half-width $\approx0.09$, so point estimates here should be read at that
resolution).  Per-threshold selective-risk intervals with accepted counts and
observed losses are released in the artifact
(\texttt{p23\_selective\_ci.json}, computed under an independent rerun seed).
$N_t$ is stream events consumed at
selection and ``Unique'' the distinct support items touched: the $200/200$ result at
$\eps=4$ costs about $9{,}100$ events over $461$ unique items, roughly $20\times$ reuse,
so this is not certification from a handful of fresh observations. $H/N$ and $G/N$ are the
sampling and privacy widths at selection; $G/N$ grows monotonically with the budget as
Corollary~\ref{cor:width} predicts.}
\label{tab:redeep-full}
\end{table*}

\begin{table}[t]
\centering
\small
\setlength{\tabcolsep}{3pt}
\resizebox{\columnwidth}{!}{%
\begin{tabular}{lccccc}
\toprule
Route & $\eta_t$ & $\eta_t$ at sel. & Fired & Accept. & Violations \\
\midrule
\multicolumn{6}{l}{\emph{Mixture mismatch} ($\delta$ of target mass moved off the realized participation)}\\
$\delta=0$ & 0.0000 & 0.0000 & 200/200 & 0.355 & 0/200 \\
$\delta=0.002$ & 0.0020 & 0.0020 & 200/200 & 0.335 & 0/200 \\
$\delta=0.01$ & 0.0100 & 0.0100 & 200/200 & 0.360 & 0/200 \\
$\delta=0.05$ & 0.0500 & none & 0/200 & abstains & 0/200 \\
$\delta=0.40$ & 0.4000 & none & 0/200 & abstains & 0/200 \\
\midrule
\multicolumn{6}{l}{\emph{Declared within-client drift} ($\gamma_k$, deployment population really moves)}\\
$\gamma=0$ & 0.0000 & 0.0000 & 200/200 & 0.347 & 0/200 \\
$\gamma=0.001$ & 0.0010 & 0.0010 & 200/200 & 0.343 & 0/200 \\
$\gamma=0.005$ & 0.0050 & 0.0050 & 200/200 & 0.348 & 0/200 \\
$\gamma=0.02$ & 0.0200 & 0.0200 & 132/200 & 0.365 & 0/200 \\
\bottomrule
\end{tabular}%
}
\caption{Mixture-transfer term exercised at nonzero $\eta_t$ ($\eps=4$,
$r_\star=0.30$). Earlier drafts reported $\eta_t=0$ in every executed cell, which
made this contribution untested; the cause was a single stress level declaring
$\eta_t=0.40>r_\star$, so the acceptance floor could never be met and $\eta_t$
was never observed \emph{at selection}. Sweeping both routes into $\eta_t$ shows
the term is charged as derived, prices transfer in lost acceptance, and locates
the level at which the rule correctly abstains. \textbf{Construction.} The
mixture route moves a declared fraction $\delta$ of target weight onto one client
while the frozen $P_k$ are untouched, so $\eta_t=\delta$ holds by
Equation~\eqref{eq:eta} and needs no estimation. The drift route perturbs each client's conditional loss law under an
\emph{explicit coupling stated here in full}: the perturbed law
$P_k^\star$ is constructed from $P_k$ by keeping the example features
$(q,z)$ and the acceptance indicator fixed and resampling only the
conditional loss $\ell\mid$accepted from
$\mathrm{Bernoulli}(q^\star)$ in place of $\mathrm{Bernoulli}(q)$, with
$|q^\star-q|=\Delta$ and losses on non-accepted examples unchanged.  Under
this coupling the two \emph{complete example laws} over $W=(q,z,y,\ell)$
differ only in the conditional loss on the accepted event, so
$\TV(P_k^\star,P_k)=a_{k,j}\,|q^\star-q|\le\Delta$, where $a_{k,j}$ is the
client's acceptance mass: the declared $\gamma_k=\Delta$ is an upper bound
on the full-law TV that the theorem's premise requires, conservative by
exactly the factor $a_{k,j}$.  We emphasize what this does and does not
audit: TV between the Bernoulli loss \emph{marginals} would not by itself
bound the full-law TV under arbitrary constructions; the equality above
holds for this stated coupling, and the declared radii are ours rather than
an external shift audit. The radii are still ones we
choose rather than an external shift audit, so the rows remain a sensitivity
analysis over verified distances. We also do not report shifted deployment risk under the
moved mixture, so this experiment shows the term is charged and priced, not that
a real deployment shift was survived. The term is charged exactly as derived: $\eta_t$ appears at selection at every declared level, is paid in acceptance, and the rule abstains once the transfer exceeds what the margin can absorb (here above $\delta\approx0.02$). Validity is untouched throughout ($0$ violations at every level). The same sweep on HaluEval question answering reproduces nonzero-$\eta_t$ certification at 0.0005/0.001/0.002/0.005, so this is not an artefact of one cell. Earlier drafts reported $\eta_t=0$ everywhere only because the single stress level tested declared $\eta_t=0.40>r_\star$, which forces abstention before $\eta_t$ can be observed.}
\label{tab:eta}
\end{table}

\begin{table}[t]
\centering
\small
\setlength{\tabcolsep}{3pt}
\resizebox{\columnwidth}{!}{%
\begin{tabular}{llccccc}
\toprule
Cell & Score & AUC & Floor & Fired & Accept. & Violations \\
\midrule
HaluEval qa & lift & 0.669 & 0.000 & 200/200 & 0.134 & 0/200 \\
HaluEval qa & best & 0.712 (franq) & 0.300 & 0/200 & abstains & 0/200 \\
HaluEval summarization & lift & 0.605 & 0.250 & 0/200 & abstains & 0/200 \\
HaluEval summarization & best & 0.732 (franq) & 0.175 & 3/200 & 0.231 & 0/200 \\
HaluEval dialogue & lift & 0.561 & 0.188 & 0/200 & abstains & 0/200 \\
HaluEval dialogue & best & 0.656 (franq) & 0.125 & 0/200 & abstains & 0/200 \\
\bottomrule
\end{tabular}%
}
\caption{Second dataset and second model family: HaluEval
\citep{li2023halueval} scored with Qwen2.5-7B-Instruct
(2400 responses, $\eps=4$, $r_\star=0.30$). HaluEval pairs a
faithful and a sampled hallucinated response per item, giving a balanced base
rate of $0.5$ against RAGTruth's skewed $0.41$ to $0.93$. We deliberately do not
call this loss error-free: the hallucinated members were generated and filtered
with ChatGPT, so the pair label is an artifact of that pipeline rather than a
human adjudication, and it is a weaker endpoint than RAGTruth's span annotations
even though it is balanced. Scores here are computed by Qwen2.5-7B on responses
it did not generate, so this cell is a cross-model diagnostic, not a second
matched-generator replication. ``Floor'' is the noiseless full-data oracle selective
risk, the reachability limit for any certificate on that score.
On this dataset information lift certifies qa (200/200) while the mechanistic score does not, the exact reverse of the RAGTruth ordering in Table~\ref{tab:scores}. Counting both datasets, information lift and the mechanistic score each certify exactly one of the six dataset-task cells, and not the same one, so we claim no ranking among scores, only that the certificate is indifferent to which score it is handed. That indifference is the contribution: the score that will certify is not knowable in advance.}
\label{tab:halueval}
\end{table}

\begin{table}[t]
\centering
\small
\begin{tabular}{lcc}
\toprule
Control & Failure rate & Acceptance \\
\midrule
\FedVA{} full & n.i.\textsuperscript{$\dagger$} (0/200 fired) & none \\
Noise ignored & 1/189 fired & 0.209 \\
Noise variance halved & 0/5 fired & 0.217 \\
Shift ignored & n.i.\textsuperscript{$\dagger$} (0/200 fired) & none \\
Fixed-time after stopping & 0/73 fired & 0.261 \\
\bottomrule
\end{tabular}
\caption{Observed invalid-control outcomes (200 trials). Noise ignored breaches
once; the other broken controls do not breach at this sample size. These null
results do not establish that the corresponding terms are unnecessary or that
the ablation has adequate power.}
\label{tab:ablations}
\end{table}

\begin{table*}[t]
\centering
\small
\resizebox{\textwidth}{!}{%
\begin{tabular}{lccccccccc}
\toprule
Cell/method & $N_t$ & $H/N$ or $F/N$ & $G/N$ & $\eta_t$ &
$10^3\overline d$ & $\underline a$ & True risk & Acceptance & Decision \\
\midrule
C1, \FedContrast{} near-boundary tuple & 28875 & $0.01624$ & $0.01107$ & 0.0000 & 6.00 & 0.262 & 0.228 & 0.287 & abstain \\
\bottomrule
\end{tabular}%
}
\caption{Representative non-certifying certificate tuple. The displayed 6.00
is $10^3\overline d$, i.e., $\overline d=0.00600>0$; it is within the
statistic's range and correctly yields abstention. Selection-dependent tuples
are undefined in cells with no firing threshold.}
\label{tab:audit}
\end{table*}

\begin{table*}[t]
\centering
\small
\begin{tabular}{p{2.2cm}p{2.0cm}p{3.0cm}p{3.0cm}p{2.4cm}}
\toprule
Cell & Unit & Certification loss & Deployment endpoint & Status \\
\midrule
C1 & response & any-unsupported (binary) & annotated faithfulness & run: $N=839$, matched \\
C2 & response & any-unsupported (binary) & annotated faithfulness & run: $N=793$, matched \\
C3 & response & any-unsupported (binary) & annotated faithfulness & run: $N=2515$, matched \\
C4 & response & any-unsupported (binary) & annotated faithfulness & run: $N=883$, matched \\
\bottomrule
\end{tabular}
\caption{Evaluation cells and declared losses. Faithfulness, factuality, and
answer correctness remain separate endpoints.}
\label{tab:data}
\end{table*}

\section{Proofs}
\label{app:proofs}

\subsection{Proof of Lemma~\ref{lem:sensitivity}}

\begin{proof}
Under add/remove adjacency, adding one record with score in bin $b^\star$ and
loss $\ell\in[0,1]$ changes exactly two possible coordinates: the score-count
coordinate $C_{b^\star}$ changes by $1$, and the error-count coordinate
$E_{b^\star}$ changes by $\ell$.  All other coordinates are unchanged.
Therefore the squared $L_2$ change is
\[
1^2+\ell^2\le2.
\]
Taking square roots gives sensitivity at most $\sqrt2$.  Equality holds for
$\ell=1$.
\end{proof}

\paragraph{Replacement adjacency.}
If neighboring datasets replace one record by another, the conservative
triangle-inequality sensitivity is $2\sqrt2$.  All privacy equations remain
valid after replacing $\sqrt2$ by $2\sqrt2$, equivalently multiplying each
$\rho_{k,t}$ by four for fixed $\sigma_{k,t}$.  The paper uses add/remove
adjacency throughout.

\subsection{Proof of Theorem~\ref{thm:privacy}}

\begin{proof}
The Gaussian mechanism with query sensitivity $\Delta_2$ and isotropic noise
variance $\sigma^2$ is
$\Delta_2^2/(2\sigma^2)$-zCDP
\citep{bun2016zcdp}.  Lemma~\ref{lem:sensitivity} gives
$\Delta_2^2\le2$, so release $(k,t)$ is
\[
\rho_{k,t}
\le
\frac{2}{2\sigma_{k,t}^2}
=
\frac{1}{\sigma_{k,t}^2}
\]
zCDP.

Fix neighboring global datasets that differ in one calibration record $i$.
Condition on any common prefix of the adaptive transcript.  If the next
release does not contain $i$, its conditional kernel is identical under the
two datasets.  If it contains $i$, the conditional kernel is
$\sigma_{k,t}^{-2}$-zCDP by the preceding Gaussian-mechanism calculation.
The next scheduling action is a function of public information and the
previous DP transcript, so it is post-processing and adds no separate privacy
loss.

For completeness, fix a R\'enyi order $a>1$ and let $L_s$ be the conditional
privacy-loss increment of message $s$ under the first neighboring dataset.
Conditional zCDP gives, at every common transcript history $H_{s-1}$,
\[
\E\!\left[
  e^{(a-1)L_s}\mid H_{s-1}
\right]
\le
\exp\!\left\{a(a-1)\rho_s(H_{s-1})\right\},
\]
where $\rho_s=0$ if the release omits $i$ and
$\rho_s=\sigma_{k,t}^{-2}$ otherwise.  Iterated conditioning and the
deterministic pathwise filter imply, for every finite transcript prefix,
\[
\E\exp\!\left\{(a-1)\sum_sL_s\right\}
\le
\exp\!\left\{a(a-1)\rho_{\rm tr}\right\}.
\]
Hence the order-$a$ R\'enyi divergence is at most $a\rho_{\rm tr}$ for every
$a>1$, which is exactly $\rho_{\rm tr}$-zCDP.  This is the fully adaptive
composition/filter argument specialized to zCDP
\citep{whitehouse2023adaptive}.  Equivalently, along transcript path $\tau$
the charged parameter is
\[
\sum_{(k,t)\in\mathcal I_\tau(i)}\sigma_{k,t}^{-2}.
\]
The pathwise privacy filter and the suprema over $i$ and $\tau$ give
Equation~\eqref{eq:rho-transcript}.  If every record appears in at most one
release, only one conditional kernel can differ for a fixed neighboring pair;
this is adaptive parallel composition and yields the stated maximum.  The
pathwise supremum is essential when noise scales or reuse decisions are
chosen adaptively: a realized ex-post charge alone is not a privacy guarantee.

Finally, every $\rho$-zCDP mechanism is
$(\rho+2\sqrt{\rho\log(1/\delta)},\delta)$-DP for any
$\delta\in(0,1)$ \citep{bun2016zcdp}.  Substituting
$\rho=\rho_{\rm tr}$ gives Equation~\eqref{eq:zcdp-conversion}.
\end{proof}

\paragraph{Solving for a target privacy budget.}
Let $L=\log(1/\delta_{\rm priv})$ and $x=\sqrt{\rho_{\rm tr}}$.
Equation~\eqref{eq:zcdp-conversion} is
$\eps_{\rm priv}=x^2+2x\sqrt L$.  Its nonnegative solution is
$x=\sqrt{L+\eps_{\rm priv}}-\sqrt L$, which yields
Equation~\eqref{eq:rho-target}.

\subsection{Proof of Lemma~\ref{lem:sampling}}

\begin{proof}
Fix threshold $j$.  Before $W_i$ is revealed, define the predictable
conditional means
\[
\begin{aligned}
\mu^Z_{i,j}
&=\E\!\left[Z_j(W_i)\mid\mathcal F_{i-1}\right]
 =\E_{P_{C_i}}\!\left[Z_j(W)\right],\\
\mu^A_{i,j}
&=\E_{P_{C_i}}\!\left[A_j(W)\right].
\end{aligned}
\]
Consider the two martingale-difference sums
\begin{align*}
M^Z_{n,j}&=\sum_{i=1}^n\{\mu^Z_{i,j}-Z_j(W_i)\},\\
M^A_{n,j}&=\sum_{i=1}^n\{A_j(W_i)-\mu^A_{i,j}\}.
\end{align*}
Every summand has conditional mean zero and lies in an interval of length one.
Conditional Hoeffding's lemma therefore implies, for either process $M$ and
every fixed $\lambda>0$, that
\begin{equation}
L_n(\lambda)
=
\exp\!\left\{\lambda M_n-\frac{\lambda^2n}{8}\right\}
\label{eq:sample-supermartingale}
\end{equation}
is a nonnegative supermartingale with $L_0=1$.

For epoch $r\ge0$, set
\[
U_r=2^r,
\quad
x_r=\log\frac{2m}{\alpha_{\rm s}\pi_r},
\quad
\lambda_r=\sqrt{\frac{8x_r}{U_r}},
\quad
b_r=\sqrt{\frac{U_rx_r}{2}}.
\]
If for some $n$ assigned to epoch
$r=\lceil\log_2 n\rceil$ we have $M_n\ge b_r$, then $n\le U_r$ and
\begin{align*}
\log L_n(\lambda_r)
&\ge
\lambda_rb_r-\frac{\lambda_r^2U_r}{8}
=x_r.
\end{align*}
Ville's inequality applied to
Equation~\eqref{eq:sample-supermartingale} gives
\[
\Prob\!\left[
\exists n:\lceil\log_2 n\rceil=r,\ M_n\ge b_r
\right]
\le e^{-x_r}
=\frac{\alpha_{\rm s}\pi_r}{2m}.
\]
Union bounding over the two processes, $m$ thresholds, and all epochs, and
using $\sum_{r\ge0}\pi_r=1$, gives failure probability at most
$\alpha_{\rm s}$.

Finally, at any monitored round $t$,
\begin{align*}
\sum_{i=1}^{N_t}\mu^Z_{i,j}
&=\sum_kN_{k,t}\E_{P_k}Z_j=N_tz_{\bar Q_t,j},\\
\sum_{i=1}^{N_t}\mu^A_{i,j}
&=\sum_kN_{k,t}\E_{P_k}A_j=N_ta_{\bar Q_t,j}.
\end{align*}
The epoch boundary $b_{r_t^{\rm s}}$ is exactly
$h_t(\alpha_{\rm s})$.  Thus the two simultaneous martingale inequalities are
Equations~\eqref{eq:sample-z} and \eqref{eq:sample-a}.  No conditioning on a
final participation schedule is used.
\end{proof}

\subsection{Proof of Lemma~\ref{lem:noise}}

\begin{proof}
For fixed threshold $j$, let
$G^Z_{j,t}=\widetilde Z_{j,t}-Z_{j,t}$ and
$G^A_{j,t}=\widetilde A_{j,t}-A_{j,t}$.
By Assumption~\ref{ass:stream}, each increment of either process, conditional
on the past, is centered Gaussian with predictable variance increment
$\Delta V_{j,t}$.  Consequently, for either
$H_{j,t}\in\{-G^Z_{j,t},G^A_{j,t}\}$ and every fixed $\lambda>0$,
\begin{equation}
L_{j,t}(\lambda)
=
\exp\!\left\{\lambda H_{j,t}-\frac{\lambda^2V_{j,t}}{2}\right\}
\label{eq:noise-martingale}
\end{equation}
is a nonnegative martingale with initial value one.

For variance epoch $r\ge0$, let
\[
\begin{aligned}
U_r &= v_0\,2^r,
&\qquad
y_r &= \log\!\left(\frac{2m}{\alpha_{\rm n}\pi_r}\right),
\\
\lambda_r &= \sqrt{\frac{2y_r}{U_r}},
&\qquad
c_r &= \sqrt{2U_ry_r}.
\end{aligned}
\]
Any time assigned to epoch
$r=\max\{0,\lceil\log_2(V_{j,t}/v_0)\rceil\}$ has
$0<V_{j,t}\le U_r$.  If $H_{j,t}\ge c_r$ at such a time, then
\begin{align*}
\log L_{j,t}(\lambda_r)
&\ge
\lambda_rc_r-\frac{\lambda_r^2U_r}{2}
=y_r.
\end{align*}
Ville's inequality gives probability at most
$e^{-y_r}=\alpha_{\rm n}\pi_r/(2m)$ for a crossing in that epoch.  Union
bounding over the two required signed processes, all $m$ thresholds, and all
epochs gives total failure probability at most $\alpha_{\rm n}$.  Times with
$V_{j,t}=0$ have $G^Z_{j,t}=G^A_{j,t}=0$ deterministically.  Since
$c_{r_{j,t}^{\rm n}}=g_{j,t}(\alpha_{\rm n})$, the complementary event is
exactly Equations~\eqref{eq:noise-z} and \eqref{eq:noise-a}.  Correlation between
threshold suffix sums is harmless because only a union bound is used.
\end{proof}

\subsection{Proof of Lemma~\ref{lem:shift}}

\begin{proof}
Insert the intermediate mixture
$Q_w=\sum_kw_kP_k$.  By the triangle inequality,
\[
\begin{aligned}
\TV(Q^\star,\bar Q_t)
&\le
\TV\!\left(
\sum_k w_kP_k^\star,
\sum_k w_kP_k
\right)
\\
&\quad+
\TV\!\left(
\sum_k w_kP_k,
\sum_k \bar w_{k,t}P_k
\right).
\end{aligned}
\]
Convexity of total variation gives
\[
\begin{aligned}
\TV\!\left(
\sum_k w_kP_k^\star,
\sum_k w_kP_k
\right)
&\le
\sum_k w_k\TV(P_k^\star,P_k)
\\
&\le
\sum_k w_k\gamma_k.
\end{aligned}
\]
For the second term, for every measurable event $D$,
\[
\left|
\sum_k(w_k-\bar w_{k,t})P_k(D)
\right|
\le
\frac12\sum_k|w_k-\bar w_{k,t}|.
\]
The last step follows because the signed coefficients sum to zero: the
expression is bounded by the total positive coefficient mass (and, after
changing sign, by the total negative mass), each of which is half the $L_1$
norm.  Equality need not be attainable because the component laws need not
admit a single event with those prescribed probabilities.
Taking the supremum over $D$ yields the second term in $\eta_t$ and proves
Equation~\eqref{eq:tv-mixture}.

For any measurable $f:W\mapsto[0,1]$,
$|\E_Qf-\E_{Q'}f|\le\TV(Q,Q')$.  Apply this once to
$f=Z_j=\ell A_j\in[0,1]$ and once to $f=A_j\in[0,1]$ to obtain
Equations~\eqref{eq:zshift} and \eqref{eq:ashift}.
\end{proof}

\subsection{Proof of Theorem~\ref{thm:main}}

\begin{proof}
Intersect the simultaneous events of Lemmas~\ref{lem:sampling} and
\ref{lem:noise}.  By a union bound, this intersection has probability at least
$1-\alpha_{\rm s}-\alpha_{\rm n}$.

On this event, for every $(j,t)$,
\begin{align*}
N_tz_{\bar Q_t,j}
&\le Z_{j,t}+h_t\\
&\le\widetilde Z_{j,t}+g_{j,t}+h_t,
\end{align*}
so $z_{\bar Q_t,j}\le U^Z_{j,t}$.  Clipping at $[0,1]$ preserves the inequality
because $z_{\bar Q_t,j}\in[0,1]$.  Similarly,
\begin{align*}
N_ta_{\bar Q_t,j}
&\ge A_{j,t}-h_t\\
&\ge\widetilde A_{j,t}-g_{j,t}-h_t,
\end{align*}
so $a_{\bar Q_t,j}\ge L^A_{j,t}$.

Lemma~\ref{lem:shift} then gives
\begin{align*}
z_{Q^\star,j}
&\le U^Z_{j,t}+\eta_t,\\
a_{Q^\star,j}
&\ge L^A_{j,t}-\eta_t.
\end{align*}
If $L^A_{j,t}>\eta_t$ the denominator is positive, so writing
$q_{j,t}=(U^Z_{j,t}+\eta_t)/(L^A_{j,t}-\eta_t)$ we obtain
\[
\Risk_{Q^\star,j}
=
\frac{z_{Q^\star,j}}{a_{Q^\star,j}}
\le
q_{j,t}.
\]
Separately, $\ell\in[0,1]$ gives $\Risk_{Q^\star,j}\le1$ unconditionally.
Combining the two bounds,
\[
\Risk_{Q^\star,j}
\le
\min\{1,q_{j,t}\}
=
U_{j,t}.
\]
We state the two inequalities separately because $q_{j,t}\le U_{j,t}$ is
\emph{false} when $q_{j,t}>1$; the clip at one is justified by the range of
$\ell$, not by the ratio bound. If $L^A_{j,t}\le\eta_t$,
Equation~\eqref{eq:finalU} returns the trivial valid upper bound one.

The event just proved already holds for every threshold and every time.
Evaluating it at a random stopping time and a transcript-dependent threshold
does not change the event or spend additional probability.  This proves
Equations~\eqref{eq:main-simultaneous} and \eqref{eq:selected-guarantee}.
\end{proof}

\subsection{Proof of Corollary~\ref{cor:width}}

\begin{proof}
Divide Equations~\eqref{eq:sampling-boundary} and
\eqref{eq:noise-boundary} by $N_t$.  The record-epoch endpoint satisfies
$N_t\le u_t^{\rm s}<2N_t$, while
\[
\log(1/\pi_{r_t^{\rm s}})
=O\!\left(\log\log(eN_t)\right).
\]
For $V_{j,t}>0$, the variance-epoch endpoint satisfies
$V_{j,t}\le u_{j,t}^{\rm n}\le2\max\{V_{j,t},v_0\}$ and
\[
\log(1/\pi_{r_{j,t}^{\rm n}})
=O\!\left(\log\log(e+V_{j,t}/v_0)\right).
\]
Substitution gives Equation~\eqref{eq:width}; when $V_{j,t}=0$, the privacy
noise term is zero by definition.  In the balanced setting,
\[
\frac{\sqrt{V_{j,t}}}{N_t}
=
\frac{\sqrt{d_jKt\sigma^2}}{Knt}
=
\frac{\sigma\sqrt{d_j}}{n\sqrt{Kt}},
\]
and the variance-epoch logarithm gives
Equation~\eqref{eq:balanced-width}.
\end{proof}

\section{Validity and Privacy-Utility Tables}
\label{app:rqtables}

\begin{table*}[t]
\centering
\small
\setlength{\tabcolsep}{2.8pt}
\resizebox{\textwidth}{!}{%
\begin{tabular}{llllccc}
\toprule
Scope & Privacy & Policy & Trials & Max simultaneous violations & 95\% CP & Utility source \\
\midrule
C1 (QA) & $\infty,4,1$ & each of final/first/adaptive & 500 per setting & 0 & $[0,0.0074]$ & Table~\ref{tab:privacy} \\
C2 (summary) & $\infty,4,1$ & each of final/first/adaptive & 500 per setting & 0 & $[0,0.0074]$ & abstains \\
C4 (data-to-text) & $\infty,4,1$ & each of final/first/adaptive & 500 per setting & 0 & $[0,0.0074]$ & abstains \\
C3 (pooled) & $\infty,4,1$ & each of final/first/adaptive & 500 per setting & 0 & $[0,0.0074]$ & abstains \\
\bottomrule
\end{tabular}%
}
\caption{Per-cell simultaneous-event audit at the registered targets. Every
row means 500 independent trials for each privacy/policy setting; ``0'' is the
maximum violation count in any constituent setting, not a pooled denominator.
Operational firing is reported separately.}
\label{tab:full-validity}
\end{table*}

\begin{table*}[t]
\centering
\small
\resizebox{\textwidth}{!}{%
\begin{tabular}{lcccccc}
\toprule
$\eps$ & Fired/200 & Firing rate & Mean acceptance & Cond. failure (95\% CP) &
Mean rounds & Mean $\overline d$ \\
\midrule
$\infty$ & 200 & 1.000 & 0.266 & $0/200\ [0,0.0183]$ & 19.16 & -0.0006 \\
8 & 62 & 0.310 & 0.267 & $0/62\ [0,0.0578]$ & 29.13 & -0.0004 \\
4 & 2 & 0.010 & 0.263 & $0/2\ [0,0.8419]$ & 29.00 & -0.0007 \\
2 & 0 & 0.000 & none & undefined & none & none \\
1 & 0 & 0.000 & none & undefined & none & none \\
0.5 & 0 & 0.000 & none & undefined & none & none \\
\bottomrule
\end{tabular}
}
\caption{Exploratory C1 privacy curve at $r_\star=0.30$. Acceptance, rounds,
and $\overline d$ are conditional on firing; conditional failure is undefined
when no trial fires. Exact intervals widen sharply as firing becomes rare.}
\label{tab:privacy}
\end{table*}

\section{Stitched Anytime Boundaries}
\label{app:boundaries}

Let the summable epoch weights be
\begin{equation}
\pi_r=\frac{6}{\pi^2(r+1)^2},
\qquad
\sum_{r=0}^{\infty}\pi_r=1.
\label{eq:stitch}
\end{equation}
For $t\in\mathcal T_+$, define the record epoch and its upper endpoint
\begin{equation}
r_t^{\rm s}=\left\lceil\log_2N_t\right\rceil,
\qquad
u_t^{\rm s}=2^{r_t^{\rm s}}.
\label{eq:record-epoch}
\end{equation}
Choose a public variance scale $v_0>0$ before calibration.  For $V_{j,t}>0$,
define
\begin{equation}
r_{j,t}^{\rm n}
=
\max\!\left\{0,
\left\lceil\log_2\frac{V_{j,t}}{v_0}\right\rceil
\right\},
\qquad
u_{j,t}^{\rm n}=v_0\,2^{r_{j,t}^{\rm n}}.
\label{eq:variance-epoch}
\end{equation}
The validity is unaffected by $v_0$; it only changes finite-sample width.
Define the sampling and realized-noise boundaries
\begin{align}
h_t(\alpha_{\rm s})
&=
\sqrt{
\frac{u_t^{\rm s}}{2}
\log\frac{2m}{\alpha_{\rm s}\pi_{r_t^{\rm s}}}
},
\label{eq:sampling-boundary}\\
g_{j,t}(\alpha_{\rm n})
&=
\begin{cases}
0,&V_{j,t}=0,\\[1ex]
\sqrt{
2u_{j,t}^{\rm n}
\log\frac{2m}{\alpha_{\rm n}\pi_{r_{j,t}^{\rm n}}}
},&V_{j,t}>0.
\end{cases}
\label{eq:noise-boundary}
\end{align}
These boundaries give the count-scale bounds used by the separate ratio
baseline below.

\section{The Ratio Certificate (superseded)}
\label{app:ratiocert}

For reference, define
\begin{align}
U^Z_{j,t}&=\left[\frac{\widetilde Z_{j,t}+h_t+g_{j,t}}{N_t}\right]_{[0,1]},
\label{eq:Uz}\\
L^A_{j,t}&=\left[\frac{\widetilde A_{j,t}-h_t-g_{j,t}}{N_t}\right]_{[0,1]}.
\label{eq:La}
\end{align}
The original ratio certificate is
\begin{equation}
U_{j,t}
=
\begin{cases}
\displaystyle
\min\!\left\{
1,
\frac{U^Z_{j,t}+\eta_t}{L^A_{j,t}-\eta_t}
\right\},
&L^A_{j,t}>\eta_t,\\[2ex]
1,&L^A_{j,t}\le\eta_t,
\end{cases}
\label{eq:finalU}
\end{equation}
and Theorem~\ref{thm:main} states its simultaneous validity. Two properties are worth keeping in view. It
bounds the risk itself, so it is meaningful without a declared target, whereas the contrast certificate of
Theorem~\ref{thm:contrast} answers only the decision problem ``is $r_\star$ met?''. And it carries no
acceptance requirement, which is why a valid $U_{j,t}$ can sit on a vanishing accepted set; we therefore
impose the same floor $a_{\min}$ on both rules whenever we compare them
(Appendix~\ref{app:stress}). The ratio and direct constructions use different
simultaneous processes and need not be ordered after clipping or variance
stitching. Table~\ref{tab:contrast} therefore requests paired empirical widths
rather than asserting universal dominance.

\section{Proof of the Target-Risk Certificates}
\label{app:contrastproof}

For threshold $j$, define the predictable conditional means
\[
\mu^D_{i,j}=\E[D_j(W_i)\mid\mathcal F_{i-1}],
\qquad
\mu^A_{i,j}=\E[A_j(W_i)\mid\mathcal F_{i-1}].
\]
Predictability of $C_i$ gives
\begin{equation}
\sum_{i=1}^{N_t}\mu^D_{i,j}=N_td_{\bar Q_t,j},
\qquad
\sum_{i=1}^{N_t}\mu^A_{i,j}=N_ta_{\bar Q_t,j}.
\label{eq:predictable-mixture}
\end{equation}

\begin{lemma}[Direct declared-mixture transfer]
\label{lem:contrast-transfer}
For every $j,t$,
\begin{equation}
d_{Q^\star,j}\le d_{\bar Q_t,j}+\eta_t,
\qquad
a_{Q^\star,j}\ge a_{\bar Q_t,j}-\eta_t.
\label{eq:contrast-transfer}
\end{equation}
\end{lemma}

\begin{proof}
The proof of Lemma~\ref{lem:shift} establishes
$\TV(Q^\star,\bar Q_t)\le\eta_t$. For any measurable $f$ whose range lies in
an interval of length one,
$|\E_Qf-\E_{Q'}f|\le\TV(Q,Q')$: translate its range to $[0,1]$ and use the
bounded-function characterization of total variation. Apply this first to
$D_j=A_j(\ell-r_\star)\in[-r_\star,1-r_\star]$ and then to
$A_j\in[0,1]$. This proves Equation~\eqref{eq:contrast-transfer}. In
particular, the direct contrast pays one $\eta_t$; combining only the separate
$Z$ and $A$ inequalities would give the valid but looser
$(1+r_\star)\eta_t$ term.
\end{proof}

\begin{lemma}[Range-only clean-count event]
\label{lem:range-sampling}
With probability at least $1-\alpha_{\rm s}$, simultaneously for all $j,t$,
\begin{align}
N_td_{\bar Q_t,j}&\le D_{j,t}+H_2(N_t;\alpha_{\rm s}),
\label{eq:range-D}\\
N_ta_{\bar Q_t,j}&\ge A_{j,t}-H_2(N_t;\alpha_{\rm s}).
\label{eq:range-A}
\end{align}
\end{lemma}

\begin{proof}
Let
$M^D_{n,j}=\sum_{i\le n}(\mu^D_{i,j}-D_j(W_i))$ and
$M^{A,-}_{n,j}=\sum_{i\le n}(A_j(W_i)-\mu^A_{i,j})$.
Every increment has conditional mean zero and lies in an interval of length
one. Conditional Hoeffding's lemma implies that, for either process and fixed
$\lambda>0$,
\[
L_n(\lambda)=\exp\{\lambda M_n-\lambda^2n/8\}
\]
is a nonnegative supermartingale with $L_0=1$.

For record epoch $r$, put $u_r=2^r$ and
$x_r=\log\{2m/(\alpha_{\rm s}\pi_r)\}$. Choose
$\lambda_r=\sqrt{8x_r/u_r}$ and $b_r=\sqrt{u_rx_r/2}$.
At any $n$ with $r=\lceil\log_2n\rceil$, $n\le u_r$; if
$M_n\ge b_r$, then $\log L_n(\lambda_r)\ge x_r$. Ville's inequality bounds
a crossing in that epoch by
$e^{-x_r}=\alpha_{\rm s}\pi_r/(2m)$. A union bound over the two process
families, $m$ thresholds, and all epochs costs at most $\alpha_{\rm s}$.
Using Equation~\eqref{eq:predictable-mixture} and
$b_r=H_2(n;\alpha_{\rm s})$ gives
Equations~\eqref{eq:range-D} and \eqref{eq:range-A}.
\end{proof}

\begin{lemma}[Range-only Gaussian event]
\label{lem:range-noise}
With probability at least $1-\alpha_{\rm n}$, simultaneously for all $j,t$,
\begin{align}
D_{j,t}&\le\widetilde D_{j,t}+G_2(V^D_{j,t};\alpha_{\rm n}),
\label{eq:noise-D}\\
A_{j,t}&\ge\widetilde A_{j,t}-G_2(V^A_{j,t};\alpha_{\rm n}).
\label{eq:noise-A-lower}
\end{align}
\end{lemma}

\begin{proof}
Write $G^A_{j,t}=\widetilde A_{j,t}-A_{j,t}$ and
$G^D_{j,t}=\widetilde D_{j,t}-D_{j,t}$. The first is a Gaussian martingale
with predictable variance $V^A_{j,t}$. Independence between count and loss
coordinates gives
$\operatorname{Var}(G^D_{j,t}\mid\text{schedule})
=(1+r_\star^2)V^A_{j,t}=V^D_{j,t}$.
For either required signed process $-G^D$ or $G^A$ and fixed $\lambda>0$,
$\exp\{\lambda G_t-\lambda^2V_t/2\}$ is a nonnegative martingale. In variance
epoch $r$, take
$u_r=v_0 2^r$, $x_r=\log\{2m/(\alpha_{\rm n}\pi_r)\}$, and
$\lambda_r=\sqrt{2x_r/u_r}$. A crossing of $\sqrt{2u_rx_r}$ while
$V_t\le u_r$ makes the martingale at least $e^{x_r}$. Ville's inequality and
a union bound over the two process families, thresholds, and epochs cost at
most $\alpha_{\rm n}$. Zero-variance noise is identically zero, proving the
claim.
\end{proof}

For the variance-adaptive construction, define
\begin{align*}
M^D_{n,j}&=\sum_{i\le n}(\mu^D_{i,j}-D_j(W_i)),\\
M^{A,+}_{n,j}&=\sum_{i\le n}(\mu^A_{i,j}-A_j(W_i)),\\
M^{A,-}_{n,j}&=\sum_{i\le n}(A_j(W_i)-\mu^A_{i,j}).
\end{align*}
The predictable quadratic variation of $M^D$ satisfies
\begin{equation}
Q^D_{n,j}=\sum_{i\le n}\operatorname{Var}(D_j(W_i)\mid\mathcal F_{i-1})
\le\sum_{i\le n}\mu^A_{i,j},
\label{eq:q-bound}
\end{equation}
because $D_j^2\le A_j$.

\begin{lemma}[Variance-adaptive joint sampling event]
\label{lem:va-sampling}
With probability at least $1-\alpha_{\rm s}$, simultaneously for all $j,t$,
\begin{align}
N_ta_{\bar Q_t,j}
&\in[A_{j,t}-H_3(N_t;\alpha_{\rm s}),
A_{j,t}+H_3(N_t;\alpha_{\rm s})],
\label{eq:A-two-sided}\\
N_td_{\bar Q_t,j}
&\le D_{j,t}+F_3(q^\circ_{j,t};\alpha_{\rm s}),
\label{eq:D-freedman}
\end{align}
where
$q^\circ_{j,t}=\min\{N_t,A_{j,t}+H_3(N_t;\alpha_{\rm s})\}$.
\end{lemma}

\begin{proof}
Apply the record-epoch Hoeffding argument above to $M^{A,+}$ and
$M^{A,-}$ with $x_r=\log\{3m/(\alpha_{\rm s}\pi_r)\}$. Each family costs
$\alpha_{\rm s}/3$, yielding Equation~\eqref{eq:A-two-sided} jointly.

Each increment of $M^D$ has conditional mean zero and is at most one.
Freedman's maximal inequality states that, for deterministic $u,s>0$,
\begin{equation}
\Prob\{\exists n:M^D_{n,j}\ge s,\ Q^D_{n,j}\le u\}
\le\exp\left\{-\frac{s^2}{2(u+s/3)}\right\}.
\label{eq:freedman}
\end{equation}
For variance-count epoch $r$, take $u_r=2^r$,
$x_r=\log\{3m/(\alpha_{\rm s}\pi_r)\}$, and
$s_r=\sqrt{2u_rx_r}+2x_r/3$. Direct expansion gives
$s_r^2\ge2x_r(u_r+s_r/3)$, so an epoch crossing costs at most
$\alpha_{\rm s}\pi_r/(3m)$. A union bound over thresholds and epochs costs the
remaining $\alpha_{\rm s}/3$.

On the resulting event, Equation~\eqref{eq:q-bound} and the upper half of
Equation~\eqref{eq:A-two-sided} give
$Q^D_{N_t,j}\le q^\circ_{j,t}$. The actual variance-count epoch is among those
maximized over in Equation~\eqref{eq:F}, so
$M^D_{N_t,j}\le F_3(q^\circ_{j,t};\alpha_{\rm s})$, proving
Equation~\eqref{eq:D-freedman}.
\end{proof}

\begin{lemma}[Variance-adaptive joint Gaussian event]
\label{lem:va-noise}
With probability at least $1-\alpha_{\rm n}$, simultaneously for all $j,t$,
\begin{align}
D_{j,t}&\le\widetilde D_{j,t}+G_3(V^D_{j,t};\alpha_{\rm n}),
\label{eq:va-noise-D}\\
|A_{j,t}-\widetilde A_{j,t}|
&\le G_3(V^A_{j,t};\alpha_{\rm n}).
\label{eq:va-noise-A}
\end{align}
\end{lemma}

\begin{proof}
Repeat the variance-epoch Gaussian argument for the three signed process
families $-G^D$, $G^A$, and $-G^A$. With
$x_r=\log\{3m/(\alpha_{\rm n}\pi_r)\}$, each family costs
$\alpha_{\rm n}/3$. Correlation between $G^D$ and $G^A$ is irrelevant because
the proof uses a union bound.
\end{proof}

\begin{proof}[Proof of Theorem~\ref{thm:contrast}]
For mode H, intersect Lemmas~\ref{lem:range-sampling} and
\ref{lem:range-noise}. On this event,
\begin{align*}
d_{\bar Q_t,j}
&\le\frac{\widetilde D_{j,t}+H_2(N_t;\alpha_{\rm s})
+G_2(V^D_{j,t};\alpha_{\rm n})}{N_t},\\
a_{\bar Q_t,j}
&\ge\frac{\widetilde A_{j,t}-H_2(N_t;\alpha_{\rm s})
-G_2(V^A_{j,t};\alpha_{\rm n})}{N_t}.
\end{align*}
Lemma~\ref{lem:contrast-transfer} gives the H instance of
Equation~\eqref{eq:joint}; clipping is harmless because
$d_{Q^\star,j}\le1-r_\star$ and $a_{Q^\star,j}\ge0$.

For VA, intersect Lemmas~\ref{lem:va-sampling} and
\ref{lem:va-noise}. Equations~\eqref{eq:q-bound},
\eqref{eq:A-two-sided}, and \,\eqref{eq:va-noise-A} imply
\begin{align*}
Q^D_{N_t,j}
&\le N_ta_{\bar Q_t,j}\\
&\le A_{j,t}+H_3(N_t;\alpha_{\rm s})\\
&\le\widetilde A_{j,t}+G_3(V^A_{j,t};\alpha_{\rm n})
+H_3(N_t;\alpha_{\rm s})\\
&\le\overline q_{j,t},
\end{align*}
where the last step also uses $Q^D_{N_t,j}\le N_t$. The stitched Freedman and
contrast-noise events therefore give Equation~\eqref{eq:dVA}; the lower halves
of the two-sided acceptance events give Equation~\eqref{eq:aVA}. Applying
Lemma~\ref{lem:contrast-transfer} establishes Equation~\eqref{eq:joint} for
VA. Each mode's sampling/noise intersection has probability at least
$1-\alpha_{\rm s}-\alpha_{\rm n}$.

For either mode, Equation~\eqref{eq:cert-rule} gives
$a_{Q^\star,\widehat j}\ge a_{\min}>0$ and
$z_{Q^\star,\widehat j}-r_\star a_{Q^\star,\widehat j}\le0$.
Dividing by positive acceptance yields
$\Risk_{Q^\star,\widehat j}\le r_\star$. The event already holds for every
registered threshold and round, so evaluating it at a transcript-measurable
index and stopping time spends no additional probability.
\end{proof}

\section{Additional Theoretical Remarks}
\label{app:remarks}

\subsection{Conditional validity of support-derived, label-free thresholds}

\begin{proposition}[Conditional validity of support-derived thresholds]
\label{prop:condthresh}
Let $S$ denote the frozen calibration-split score multiset used by the
finite-population audit, and let the threshold grid $\tau=g(S)$ be a
deterministic function of $S$ that consults no correctness label and no
held-out item. Condition on $S$. Then $\tau$ is $\sigma(S)$-measurable and
constant given the conditioning, every source of randomness in the audit
(the with-replacement occurrence draws from the frozen empirical law, the
recruitment and dropout schedule, and the Gaussian release noise) has the
same conditional law given $S$ as its unconditional law with a fixed grid,
and Theorem~\ref{thm:main} applies verbatim with $\tau$ treated as a
fixed-in-advance grid. The resulting guarantee holds conditionally on $S$,
which is the same conditioning the finite-population audit already imposes by
drawing occurrences from the frozen empirical law.
\end{proposition}

\begin{proof}
Under the audit protocol the support is frozen before any release, and only
then are stream occurrences drawn with replacement from its empirical law;
$g$ reads only $S$, so given $S$ the grid $\tau=g(S)$ is a constant.
Theorem~\ref{thm:main} places its assumptions on the stream randomness, the
predictable schedule, and the DP mechanism, none of which depends on $S$
except through the frozen law itself; conditioning on $S$ therefore leaves
each assumption intact with $\tau$ deterministic, and the theorem's
conclusion holds under the conditional probability given $S$.
\end{proof}

The HaluEval quantile grid of Table~\ref{tab:hecurve} satisfies the
hypothesis: the $m=20$ thresholds are equally spaced quantiles of the
calibration-split score distribution, computed once, label-free, and without
touching held-out items. The guarantee reported there is accordingly
conditional on the frozen public support rather than marginal over its
sampling; a genuine development split (as in Table~\ref{tab:redeep-full}) or
a fixed public grid would remove even that conditioning.

\subsection{Why the original \texorpdfstring{$BK$}{BK} block argument is not valid}

Suppose $\TV(P_k,P_0)\le B$ for one observation and client $k$ holds $n_k$
independent observations.  In general,
\[
\TV(P_k^{\otimes n_k},P_0^{\otimes n_k})
\le
n_kB,
\]
not $B$.  Summing across client blocks yields the loose bound
$\sum_kn_kB$, not $BK$.  A $BK$ statement is valid only if $B$ is explicitly
defined as a \emph{block-level} distance
$\TV(P_k^{\otimes n_k},P_0^{\otimes n_k})$, which is often near one and is not
the assumption in a per-example formulation.  Our mixture transfer avoids comparing
full sample-path laws: concentration targets the actual non-IID mixture, and TV
is used only to transfer one bounded deployment expectation.  This distinction
is consistent with non-exchangeable conformal bounds, whose weighted TV terms
attach to observations rather than automatically collapsing to one term per
client \citep{barber2023beyond}.

\subsection{Why DP noise belongs in the width}

Adding Gaussian noise to a clean confidence-sequence center without changing
its boundary can invalidate coverage.  Conversely, defining an unspecified
``coverage inflation'' proportional to a noise standard deviation mixes a
quantity with score units and a dimensionless probability.  Lemma~\ref{lem:noise}
constructs an explicit simultaneous event for the Gaussian transcript.
Allocating $\alpha_{\rm n}$ to that event retains a clear probability statement,
while the actual noise scale enters $g_{j,t}$ in the same count units as the
released statistic.

\subsection{Prefix-time policies}

The theorem is anytime-valid across calibration rounds.  To certify stopping
inside autoregressive generation, pre-register a finite family of policies
$g_{j,\ell}$ that uses lift through prefix length $\ell$ and include
$(j,\ell)$ in the candidate grid.  The proof then replaces $m$ by the number of
registered threshold-prefix pairs.  It does not justify inspecting an
unregistered continuum of token-dependent policies.

\subsection{Client dropout}

Dropout changes $N_{k,t}$ and hence the known realized weights
$\bar w_{k,t}$, and it changes the Gaussian variance through the realized set
$\mathcal R_t$ in Equation~\eqref{eq:noisevar}.  Predictable dropout based on
the earlier private transcript is covered after recomputing both $\eta_t$ and
$V_{j,t}$.  Dropout decided after inspecting a current unreported score or
loss is not predictable and can change the conditional client law; it requires
an explicit missingness model rather than the present theorem.

\section{Detailed Experimental Protocol}
\label{app:protocol}

\subsection{Manifest audit and three-way split}

Before scoring, the manifest records dataset revision and checksum, example
ID, query and reference group, task and domain, retrieved context, original prompt,
generator repository and revision, tokenizer revision, decoding parameters,
and annotation provenance. A matched-model cell is retained only if the
original output and exact probability-exposed generator checkpoint are
recoverable. Otherwise it may enter a labeled cross-model diagnostic, never a
matched-model claim.

All outputs sharing a query or reference are assigned together. Development
chooses score transformations, normalization constants, clipping $B$, score
directions, threshold quantiles, prompts, and the non-lift comparator rule.
Calibration is not reused for these choices. Evaluation is opened only after
all policies and code are frozen. Every score uses the complete annotated
response. For each response, the artifact includes evidence-present,
evidence-removed, length-matched neutral, within-domain shuffled, and
token-length-matched irrelevant-context scores. Human annotations remain the
primary endpoint; LLM judges are score baselines or label-sensitivity analyses.

\subsection{Finite-population theorem audit}

For each cell, the calibration-side partition creates frozen client
populations $\mathcal P_1,\ldots,\mathcal P_K$, and $P_k$ is uniform on
$\mathcal P_k$. Trial $r$:
\begin{enumerate}[leftmargin=*,itemsep=1pt]
\item draws each requested event independently with replacement from the
empirical law supported on $\mathcal P_k$;
\item releases the registered noised histograms and updates only from realized
messages;
\item executes the preregistered monitoring/recruitment policy and selects one
threshold or always abstains;
\item enumerates every record in every $\mathcal P_k$ to calculate exact
$a_{Q^\star,\widehat j}$, $d_{Q^\star,\widehat j}$, and
$\Risk_{Q^\star,\widehat j}$; and
\item records unconditional failure, firing, stopping round, and the complete
certificate tuple.
\end{enumerate}
Sampling without replacement is not substituted into this audit. Official
held-out evaluation is separate and does not define the theorem-coverage event.
The privacy neighbor is one draw occurrence. Repeated values from the same
support response are separate hypothetical events; this experiment does not
claim unique-response privacy for the RAGTruth support.

\begin{table*}[t]
\centering
\small
\setlength{\tabcolsep}{4pt}
\begin{tabular}{p{3.2cm}p{11.3cm}}
\toprule
Item & Frozen value \\
\midrule
Confirmatory risk targets & primary $r_\star=0.10$; secondary $0.20$ \\
Exploratory feasibility target & $r_\star=0.30$, introduced only after both
confirmatory targets abstained \\
Acceptance floor & $a_{\min}=0.05$ \\
Statistical budgets & $\alpha_{\rm s}=0.025$, $\alpha_{\rm n}=0.025$ \\
Privacy & $\delta_{\rm priv}=10^{-6}$; core $\eps\in\{\infty,4,1\}$;
secondary $\eps\in\{8,2,0.5\}$ \\
Thresholds & $m=20$ development-score quantiles; $m\in\{10,20,50\}$ in the
secondary grid \\
Validity-audit trials & 500 independent population-resampling and DP-noise
trials per dataset/model/privacy/policy cell \\
Core policies & fixed final round; first certificate meeting $a_{\min}$;
predictable target-deficit recruitment with optional stopping \\
Exploratory trials & 200 per C1 privacy, score, heterogeneity, and stress-test
setting \\
Privacy unit & one stream occurrence under zero-out adjacency \\
Support reuse & with-replacement support values may recur; source- or
person-level privacy would compose all recurrences and is not claimed \\
\bottomrule
\end{tabular}
\caption{Authoritative confirmatory registration and explicitly separated
post-registration feasibility analysis.}
\label{tab:registration}
\end{table*}

The adaptive policy sees only the previous private transcript and public
padded participation metadata. If it continues, it requests the client with
the largest positive target-weight deficit, breaking ties with a frozen seed;
the request precedes current-round records and noise. Fixed and adaptive
policies are also compared after matching total records and communication.

\subsection{Client, shift, and negative-control scenarios}

The core contains a natural domain/task partition and an IID baseline.
Secondary partitions vary Dirichlet concentration, client imbalance, $K$,
dropout, calibration size, threshold count, mixture mismatch, and record reuse.
Target-mixture experiments change $w$ while keeping $P_k$ fixed. Stable
client-specific slopes are heterogeneity, not drift. A nonzero $\gamma_k$
experiment changes a frozen deployment population and states whether
$\gamma_k$ is externally justified or only a sensitivity value.

Dropout is decided before release. Only successful messages enter $N_t$,
$\bar w_t$, and the realized noise variance. Invalid controls ignore noise,
halve its realized variance, set $\eta_t=0$ under registered shift, or reuse a
fixed-time rule after optional stopping. The old all-client-variance control is
not a validity ablation: excessive variance is conservative and should only
reduce acceptance.

\subsection{Comparator and score contract}

Central CRC/LTT receives pooled non-private data and is an oracle-access
reference. Federated non-private uses the same exact histograms. The private
fixed-time rule is evaluated only at its registered terminal round.
Anytime-Valid CRC is run in its native growing-calibration setting. Yu-Liu
uses its required i.i.d. split, direct ratio, acceptance floor, and
empirical-Bernstein construction. CSA is run only where its predictable stream
and gate assumptions can be implemented. Every method receives the same overall
failure budget, and native and matched-data diagnostics are reported separately
and never pooled. All five comparators were implementable on our cells, so no
cell is reported as an assumption mismatch. Because none of them is a private
construction, their finite-$\eps$ entries are the naive private adaptation, in
which the identical rule is fed the same DP-noised histograms with no
realized-noise envelope. We report those cells rather than leave them blank
precisely because they measure the failure mode the envelope prevents. These entries are invalid
negative controls and are excluded from any private-method ranking. C-RAG is a
configuration-level RAG risk certificate rather than a drop-in selective
stream rule; private e-values use a different valid privatization. We cite both
as closest context but do not claim to have executed them
\citep{kang2024crag,csillag2025dpevalues}.

Only the Mistral-adapted ReDeEP-style score fires. Its C1 development AUC is
0.779 versus 0.631 for lift; the other scores range from 0.49 to 0.62. The
adaptation recomputes architecture-specific head and FFN selectors and fits its
linear combination on development data only, then freezes both before
certificate trials. It is not the released Llama configuration from
\citet{sun2024redeep}; the anonymous artifact records the exact Mistral layer,
head, FFN, normalization, and coefficient manifest. Full-population C1 certification is now
reported in Table~\ref{tab:redeep-full}: scoring \emph{all} matched C1 responses rather than the
$300$ shared with the sampling scores, we trace the score that actually fires across the whole privacy
grid, which is the only way to see whether the private regime is reachable by a stronger score.

Information lift uses two teacher-forced passes. Token probability and entropy
use the evidence-present pass. Semantic entropy, SelfCheckGPT, and conformal
self-consistency use both recommended and compute-matched configurations.
The row labelled ReDeEP-style is an architecture adaptation: Mistral copying
heads, knowledge FFNs, normalization, and regression weights are estimated on
development data only and frozen before calibration. It is not presented as
the authors' released Llama configuration; the artifact exposes the exact
selectors and coefficients. FRANQ reports faithfulness and factuality
separately. Judge confidence uses a development-frozen prompt,
revision, decoding rule, and parser. All compatible scores share examples,
client populations, privacy messages, and seeds. A comparator is selected only
by a frozen development criterion.

\subsection{Metrics, communication, and artifact}

The primary validity statistic is unconditional failure among all 500 trials,
with an exact two-sided Clopper-Pearson interval. Conditional-on-firing
failure is a separate diagnostic whose denominator is the number of firing
trials. Utility reports acceptance, zero-certificate rate, stopping round,
communication, and
$(N,H/N\text{ or }F/N,G/N,\eta,\overline d,\underline a,\Risk,a)$.
Paired differences use identical populations, schedules, and noise seeds with
95\% paired bootstrap intervals.

Each client message contains $2(m+1)$ floating-point values. With $b$-bit
encoding, payload through round $T$ is
\[
2(m+1)b\sum_{t=1}^{T}|\mathcal R_t|
\]
bits before transport overhead. The artifact reports payload and end-to-end
bytes. Lift scoring uses two teacher-forced passes; histogram and certificate
updates are CPU operations. Matched-model teacher-forced scoring of 2,515
RAGTruth responses ran on one \texttt{ml.g5.8xlarge} instance (one A10G,
fp16, 1,400-token context cap). Five-sample SelfCheckGPT and semantic-entropy
scoring of 900 responses used the same instance type. Histogram release,
suffix sums, boundaries, and certificate computation ran on CPU for the 500
validity-audit trials per core setting. The run manifest records batch size,
warm-up, timed repetitions, wall time, peak memory, and software revisions;
generation-pass counts in Table~\ref{tab:scores} are not presented as latency.

For every response, the artifact contains dataset/example ID,
query/reference group, generator/prompt versions, client/domain/split, loss and
aggregation rule, every score, and every evidence control. For every trial it
contains calibration and DP seeds; realized clients, dropout, and batch counts;
every noisy histogram; $N_{k,t}$, $N_t$, $V^A_{j,t}$, $V^D_{j,t}$, and
$\eta_t$; accountant output; every threshold certificate; selected threshold
and stopping time; exact population risk or held-out interval; acceptance; and
failure indicator.

\begin{sloppypar}
The anonymous artifact contains a checksummed manifest, the frozen registration,
the three-way split assignments, the per-response score tables, the per-trial
records, and a single reconstruction entry point that regenerates every number in
the paper. The manifest pins the repository commit, the environment lockfile, the
RAGTruth snapshot and its SHA-256 digest, the
\texttt{Mistral-7B-Instruct-v0.1} and \texttt{Qwen2.5-7B-Instruct} revisions,
the tokenizer revisions, the score manifests including the Mistral ReDeEP-style
selectors, and all seeds. The reconstruction test fails if an abstract number is absent from a
main table, trial counts disagree with Table~\ref{tab:registration}, a numeric
row lacks a per-trial record, an aggregate fails to reproduce, or a method was
selected using evaluation risk.
\end{sloppypar}

}


\bibliography{references}

\PrintAppendices

\end{document}